\documentclass[11pt,a4paper]{article}
\usepackage{jheppub}
\usepackage{amsmath}
\usepackage{multicol,bbm}
\usepackage{enumerate}
\usepackage{bibentry}

\graphicspath{{./Figs/}}

\def\beq{\begin{equation}}
\def\eeq{\end{equation}}

\def\be{\begin{equation}}
\def\ee{\end{equation}}
\def\ba{\begin{eqnarray}}
\def\ea{\end{eqnarray}}
\def\bea{\begin{eqnarray}}
\def\eea{\end{eqnarray}}
\def\eq{\begin{equation}}
\def\eqe{\end{equation}}
\def\eqa{\begin{eqnarray}}
\def\eqae{\end{eqnarray}}

\def\beqa{\begin{eqnarray}}
\def\eeqa{\end{eqnarray}}

\def\e{\epsilon}

\def\de{\delta}

\newcommand{\braket}[2]{\langle #1  #2 \rangle}
\newcommand{\beqas}{\begin{eqnarray*}}

\newcommand{\eeqas}{\end{eqnarray*}}

\title{From U(1) to E$\mathbf{_8}$: soft theorems in supergravity amplitudes }
\author[a]{Wei-Ming Chen}
\author[a,b]{ Yu-tin Huang}
\author[c]{Congkao Wen}
\affiliation[a]{Department of Physics and Astronomy, National Taiwan University, Taipei 10617, Taiwan, ROC} 
\affiliation[b]{School of Natural Sciences, Institute for Advanced
Study, Princeton, NJ 08540, USA}
\affiliation[c]{Dipartimento di Fisica, Universit\`a di Roma ``Tor Vergata" \& I.N.F.N. Sezione di Roma ``Tor Vergata", Via della Ricerca Scientifica, 00133 Roma, Italy}
\emailAdd{tainist@gmail.com,yutinyt@gmail.com,Congkao.Wen@roma2.infn.it} 

\abstract{It is known that for $\mathcal{N}=8$ supergravity, the double-soft-scalar limit of an $n$-point amplitude is given by a sum of local SU(8) rotations acting on an $(n{-}2)$-point amplitude. For $\mathcal{N}<8$ supergravity theories, complication arises due to the presence of a U(1) in the U($\mathcal{N}$) isotropy group, which introduces a soft-graviton singularity that obscures the action of the duality symmetry. In this paper, we introduce an anti-symmetrised extraction procedure that exposes the full duality group. We illustrate this procedure for tree-level amplitudes in $4\leq\mathcal{N}<8$ supergravity in four dimensions, as well as $\mathcal{N}=16$ supergravity in three dimensions. In three dimensions, as all bosonic degrees of freedom transform under the E$_8$ duality group, supersymmetry ensures that the amplitude vanishes in the single-soft limit of all particle species, in contrast to its higher dimensional siblings. Using recursive formulas and generalized unitarity cuts in three dimensions, we demonstrate the action of the duality group for any tree-level and one-loop amplitudes. Finally we discuss the implications of the duality symmetry on possible counter terms for this theory. As a preliminary application, we show that the vanishing of single-soft limits of arbitrary component fields in three-dimensional supergravity rules out the direct dimensional reduction of $D^8R^4$ as a valid counter term. }
\preprint{ROM2F/2014/09}

\begin{document}
\maketitle 
\section{Introduction and motivations}
Scattering amplitudes often exhibit universal behaviors in the limit when the momenta of some external particles approach to zero, i.e. so-called soft limit. For instance, it is well known that amplitudes in gauge theories (and gravity) behave universally in the single soft gluon (and graviton) limit, which goes back to the classical work by Weinberg~\cite{Weinberg}. In particular, the analytic behavior of this limit at tree-level is completely determined by the gauge symmetries of the theory~\cite{Low, SoftGravy1, BernGauge}.\footnote{For the understanding on soft behaviors from other symmetry principles see~\cite{CS, SymArg}}

Another famous and well-studied case of soft limit, which will be of our interest in this paper, is the soft-pion theorem. The theorem states that the Goldstone boson decouples at zero momentum, i.e.  the amplitude of one soft ``pion" with arbitrary number of hard ``pions" vanishes~\cite{AdlerZero}. The full algebra of the symmetry can be exposed by considering the limit where two Goldstone bosons become soft~\cite{Weinberg:1966kf}. This idea of probing the global symmetries of the theory by studying the single- and double-soft scalar limits was revisited and applied to $\mathcal{N}=8$ supergravity theory in four dimensions by Arkani-Hamed et al~\cite{Simplest}. It is known that the theory contains $70$ scalars, which are elements in the coset space E$_7$/SU(8), thus according to the soft-pion theorem, the amplitudes vanish in the single-soft-scalar limit, which is indeed the case as shown in~\cite{Simplest}. The authors of ref.~\cite{Simplest} then beautifully showed that any $n$-point amplitude in the double-soft-scalar limit has the following universal behavior:
\eq
M_n\left( \phi^{I I_1 I_2 I_3}(  \epsilon^2 p_1),\; \phi_{J I_1 I_2 I_3}( \e^2 p_2),3,\cdots,n \right)\bigg|_{\epsilon \rightarrow 0}=
{1 \over 2}\sum_{a=3}^{n} \frac{p_a\cdot (p_1-p_2)}{p_a\cdot(p_1+p_2)}(R_{a})^I\,_J M_{n-2}+\mathcal{O}(\e)\,,
\eqe
where the superscripts in scalar field $\phi$ are the SU(8) R-symmetry indices, and $(R_{a})^I\,_J$ is the corresponding SU(8) rotation. It might be a surprise that amplitudes vanish in the single-soft limit, but finite in the double-soft limit. As explained in ref.~\cite{Simplest},  which we will give a brief review in the next section, this is a reflection of the fact that the commutators of the broken generators do not vanish. 

For $4\leq\mathcal{N}<8$ supergravity, one can proceed and derive the corresponding behavior via supersymmetry reduction of the $\mathcal{N}=8$ theory. There is one caveat however, in that for $\mathcal{N}<8$, the isotropy group (the $H$ of coset $G/H$) is U$(\mathcal{N})$ which includes a U(1). In order to generate this U(1) factor, the scalars chosen for the double-soft limit form an SU($\mathcal{N}$) singlet, which is known to be polluted by the singularity from an internal soft graviton.

In this paper, to extract the U(1) part of the duality group and subtract the singularity, we take the double-soft limit in a manifest anti-symmetric fashion with respect to the two scalars. More precisely, we consider the difference of two distinct amplitudes, one with the ($\phi_{T}$, $\phi_{\bar{T}}$) scalars carrying momenta $(p_1,p_2)$, the other with $\phi_{T}$ and $\phi_{\bar{T}}$ exchanged. We will show that 
\eqa\label{Predict}
\nonumber (\mathcal{N}=4)\;&&\bigg[M_n\left( \phi( \e^2 p_1),\; \bar{\phi}( \e^2 p_2),3\cdots,n \right)-M_n\left( \bar{\phi}( \e^2 p_1),\; \phi(\e^2 p_2),3\cdots,n \right)\bigg]\bigg|_{\e\rightarrow0}\\
\nonumber&&=\sum_{a=3}^{n} \frac{p_a\cdot (p_1-p_2)}{2p_a\cdot(p_1+p_2)}\left(R_{a}\right) M_{n-2}+\mathcal{O}(\e)\,,\\
\nonumber (\mathcal{N}=5)\;&&\bigg[M_n\left( \phi^I( \e^2 p_1),\; \bar{\phi}_I( \e^2 p_2),3\cdots,n \right)-M_n\left( \bar{\phi}_I( \e^2 p_1),\; \phi^I(\e^2 p_2),3\cdots,n \right)\bigg]\bigg|_{\e\rightarrow0}\\
\nonumber&&=\sum_{a=3}^{n} \frac{p_a\cdot (p_1-p_2)}{2p_a\cdot(p_1+p_2)}\left(\left(R_a\right)^I_I+\frac{\mathcal{N}-8}{2\mathcal{N}}\delta^{I}_{I}R_{a}\right) M_{n-2}+\mathcal{O}(\e)\,,\\
\nonumber (\mathcal{N}=6)\;&&\bigg[M_n\left( \phi^{IJ}( \e^2 p_1),\; \bar{\phi}_{IJ}( \e^2 p_2),3\cdots,n \right)-M_n\left( \bar{\phi}_{IJ}( \e^2 p_1),\; \phi^{IJ}(\e^2 p_2),3\cdots,n \right)\bigg]\bigg|_{\e\rightarrow0}\\
&&=\sum_{a=3}^{n} \frac{p_a\cdot (p_1-p_2)}{2p_a\cdot(p_1+p_2)}\left(\left(R_a\right)^I_I\delta^{J}_{J}+\left(R_a\right)^J_J\delta^{I}_{I}+\frac{\mathcal{N}-8}{2\mathcal{N}}\delta^{IJ}_{IJ}R_{a}\right) M_{n-2}+\mathcal{O}(\e)\,,
\eqae
where $R_{a}$ is the single site U(1) generator and $(R_a)^{I}\,_{I}$ is the diagonal component of the SU($\mathcal{N}$) generator $(R_a)^{I}\,_{J}\equiv \eta_a^I\frac{\partial}{\partial \eta_a^J}-\frac{\delta^I_J}{\mathcal{N}} \sum_{l=1}^{\mathcal{N}}\eta_a^l\frac{\partial}{\partial \eta_a^l}$. We will refer to such extraction of the double-soft limit as ``anti-symmetrized extraction". Note that due to the fact that we are considering non-maximal supergravity theories, the on-shell degrees of freedom are carried by two distinct multiplets ($\Phi^{\mathcal{N}},\overline{\Phi}^{\mathcal{N}}$). As a result, the U(1) generator $R_a$ has a different constant for the two distinct multiplets
\eq\label{U1Def}
R_{a}=\sum_{I}\eta_a^{I}\frac{\partial}{\partial \eta_a^I}\quad (a\in \Phi^{\mathcal{N}})\,, \quad\quad R_{a}=\sum_{I}\eta_a^{I}\frac{\partial}{\partial \eta_a^I}-\mathcal{N} \quad (a\in \overline{\Phi}^{\mathcal{N}})\,.
\eqe

We also consider maximal supergravity in three dimensions, which is the $\mathcal{N}=16$ theory introduced by Marcus and Schwarz~\cite{E8}. The 128 bosonic states now parametrize the coset E$_{8(8)}$/SO(16). We use the three-dimensional recursion formulas~\cite{3DRecur}, to derive the double-soft-scalar limit for all multiplicity tree-level amplitudes. Since the on-shell superspace only manifests U(8) $\in$ SO(16), the other part of the SO(16) generators are non-linearly realized. Thus using the double-soft limit allows us to construct the algebra of $E_{8(8)}$ in such non-linear realization. Note that the presence of a U(1) again requires us to apply the anti-symmetrized extraction procedure discussed above. We also consider the fate of the duality at loop-level. We demonstrate that at one loop, in the scalar integral basis representation, the integral coefficients are given in such a way that the double-soft behavior is manifest. 

One of the many important questions one can ask for a gravitational S-matrix is its ultraviolet behavior. In recent years tremendous progress in computation techniques has allowed us to peer ever deeper into perturbative gravitational S-matrix. Remarkably, explicit computations~\cite{N8UV, N4UV, N5UV} have reveal surprising finiteness in a wide range of supergravity theories with $4\leq\mathcal{N}\leq8$. Although from the viewpoint of four-dimensional divergences, some results can be explained by the constraints imposed by the symmetries of the coset space~\cite{ElvangR4, ElvangFull}, there are examples where finiteness requires explanations that go beyond that explained by traditional symmetry arguments~\cite{N4UV, N5UV, BCJEvid}. 

If four-dimensional maximal supergravity is finite, then so must its three-dimensional reduction. Unlike in four dimensions, here \textit{all} bosonic degrees of freedom transform under the duality group, which implies that coset symmetry imposes stronger constraints on candidate ultraviolet (UV) counter terms. Furthermore, as we will demonstrate, supersymmetric Ward-identities require that amplitudes vanish as well in the fermionic single-soft limits. Thus one can ask whether or not candidate UV counter terms can produce matrix elements satisfying all single- and double-soft behaviors required by the symmetries. As a preliminary step, we consider the direct dimensional reduction of matrix elements of $D^nR^4$ counter terms in four dimensions. We will explicitly show that these matrix elements, which satisfy the E$_7$ duality symmetry in four dimensions~\cite{ElvangFull}, do not have the correct single-soft behavior in three dimensions.


This paper is organized as following: In the next section, we study the double-soft-scalar limit for four-dimensional $\mathcal{N}=4,5,6$ supergravity theories. These theories can be studied from the $\mathcal{N}=8$ theory via supersymmetry reduction. However unlike their ancestor, the isotropy group of the duality symmetries for these non-maximal supersymmetric theories contain a U$(1)$ factor. To extract this subtle contribution, we introduce a procedure ``anti-symmetrised extraction", which allows us to throw away unwanted singular parts, and leave behind a beautiful and finite result, corresponding precisely to the U$(1)$ factor. In section \ref{section:3D}, we then move on to study $\mathcal{N} = 16 $ supergravity in three dimensions, both at tree and loop level. At tree level, we study the soft limits using BCFW recursion relations in three dimensions, and the same ``anti-symmetrised extraction" procedure introduced previously is used to extract the U$(1)$ factor in the symmetry group. After deriving the soft theorems for tree-level amplitudes, we study the possible loop corrections to the theorems. Using generalized unitarity cuts, we show that all one-loop amplitudes satisfy exactly the same soft theorems as the tree-level one. In section \ref{section:counterterm}, we discuss the application of duality symmetry to constrain candidate counter terms for $\mathcal{N} = 16 $ Supergravity. We show that S-matrix generated by many counter terms descendant from four-dimensional ones via direct dimensional reduction do not satisfy the single-soft-scalar theorems. Finally in section \ref{section:conclusion}, we finish the paper with conclusions and remarks.

\section{Soft-scalar limits in $\mathcal{N}\leq8$ Supergravity} \label{section:4D}
\subsection{Review on single- and double-soft limits on gravity amplitudes}
Massless scalars that can be identified as goldstone bosons of spontaneous broken symmetry, exhibit simple behavior in the soft limits. For theories that involve these massless scalars, in the limit where the momentum of one of these scalars becomes soft the corresponding amplitude vanishes, a result that is famously known as ``Adler's zero"~\cite{AdlerZero}. Consider the coset space $G/H$, where the generators of the isometry group $G$ are represented by $(T_i, H_j)$, and $H_i$'s are the elements of the isotropy group. Schematically they satisfy the following commutation relations:
\eq
[T,T]\sim H\,, \quad [T,H]\sim T\,, \quad [H,H]\sim H\,.
\eqe
Since the vacuum expectation values (vev) of scalars spontaneously break the symmetry, thus they can be identified with parameters of the broken generators $T_i$.  The vanishing of the soft-scalar limit can be understood through the fact that for the non-linear sigma model, which is the effective action for the goldstone bosons, scalar interactions are constructed out of covariant derivatives 
\eq
P_\mu=(e^{\varphi}\partial_\mu e^{-\varphi}-e^{-\varphi}\partial_\mu e^{\varphi})\,,
\eqe   
where $\varphi=\phi^iT_i\,.$ Since the scalars are dressed with derivatives, taking the momentum soft results in the vanishing of the amplitude. 

In~\cite{Simplest}, the soft-scalar limits were discussed without relying on any detailed structure of the interactions. Starting with the fact that spontaneous symmetry breaking is a reflection of the presence of continuous set of degenerate vacua, perturbative amplitudes computed at different points on this moduli space must be equivalent. As two different points in the moduli space are connected via the generators $T_i$, we can schematically write 
\eq\label{Vac}
|\theta+\Delta\theta\rangle=e^{i \Delta\theta\cdot T}|\theta\rangle\,,
\eqe 
where $\theta$ represents the vev of the scalar, which parametrizes the vacuum. Assuming that each point in the moduli space can be connected in such fashion, the fact that amplitudes computed in distinct vacua must agree implies that as one expands the exponent in eq.(\ref{Vac}), terms beyond the leading term in the expansion must vanish. Since $\Delta\theta$ simply corresponds to a constant scalar, i.e. scalars with zero momenta, this leads to the conclusion that amplitudes with any additional soft scalar must vanish.

However, as discussed in~\cite{Simplest}, the above analysis is not entirely correct. The subtlety lies in the assumption that there is a well-defined path that connects two points. Indeed, one would expect that the difference between two different paths should be proportional to an $H$ generator, since $[T,T]\sim H$. In~\cite{Simplest} it was argued that this ambiguity leads to the result that in the double-soft-scalar limit, the amplitude is non-vanishing, and behave universally: 
\eq
M_n\left( \phi^i( \e^2 p_1),\; \phi^j(\e^2 p_2),3\cdots,n \right)\bigg|_{\e \rightarrow0}=
{1 \over 2} \sum_{a=3}^{n} \frac{p_a\cdot (p_1-p_2)}{p_a\cdot(p_1+p_2)}[T_i,T_j]_a M_{n-2} \,.
\eqe
For $\mathcal{N}=8$ supergravity, whose 70 scalars parametrized E$_7$/SU(8) coset, this becomes 
\eq
\hspace{-0.4cm}M_n\left( \phi^{I I_1 I_2 I_3}(  \epsilon^2 p_1),\; \phi_{J I_1 I_2 I_3}( \e^2 p_2),3,\cdots,n \right)\bigg|_{\epsilon \rightarrow 0}=
{1 \over 2}\sum_{a=3}^{n} \frac{p_a\cdot (p_1-p_2)}{p_a\cdot(p_1+p_2)}(R_{a})^I\,_J M_{n-2}\,,
\eqe
where $(R_{a})^I\,_J=\eta^I_a\frac{\partial}{\partial \eta^J_a}$ is the single-site SU(8) R-symmetry generator.

For $\mathcal{N}<7$ supergravity, the scalars parametrize SO(12)/U(6), SU(1,5)/U(5) and SU(1,1)/U(1) cosets for $\mathcal{N}=6,5$ and $4$ respectively. One would expect the double-soft limits for these theories should directly follow from the $\mathcal{N}=8$ theory via SUSY reduction. This would indeed be the case if not for the subtle difference between the isotropy group for the $\mathcal{N}<7$ theories from the maximal theory: they contain an extra U(1). To see the U(1) in the double-soft limit, the two scalars must form an SU($\mathcal{N}$) singlet, which induces singularities in the limit. More precisely, the duality group algebra now involves relations of the form:
$$[T, \bar{T} ]\sim U(1)$$
where $T$ and $\bar{T}$ have opposite charges under the U(1). This implies that the double-soft-limit for such scalars will involve Feynman diagrams where the two-scalars merge into a graviton:
$$\includegraphics[scale=0.7]{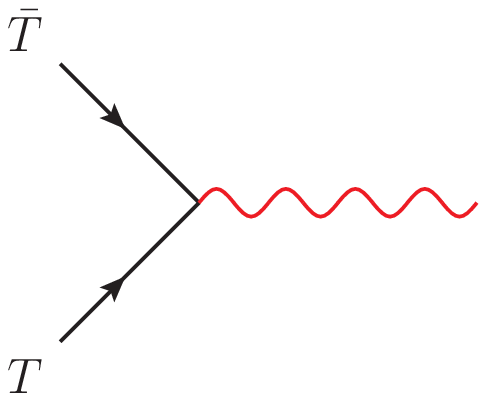} ~.$$
As the graviton is soft, the amplitude is then proportional to the soft-graviton limit of an $(n{-}1)$-point amplitude which is divergent.

To extract the U(1) part of the duality group, we take the double-soft limit in a manifest anti-symmetric fashion with respect to the two scalars. More precisely, we consider the difference of two distinct amplitudes, one with the ($\phi_{T}$, $\phi_{\bar{T}}$) scalars carrying momenta $(p_1,p_2)$, the other with $\phi_{T}$ and $\phi_{\bar{T}}$ exchanged. For $\mathcal{N}=4, 5,6$ this corresponds to considering the following difference
\eqa
\nonumber (\mathcal{N}=4)\;&&\bigg[M_n\left( \phi( \e^2 p_1),\; \bar{\phi}( \e^2 p_2),3\cdots,n \right)-M_n\left( \bar{\phi}( \e^2 p_1),\; \phi(\e^2 p_2),3\cdots,n \right)\bigg]\bigg|_{\e\rightarrow0}\,,\\
\nonumber (\mathcal{N}=5)\;&&\bigg[M_n\left( \phi^I( \e^2 p_1),\; \bar{\phi}_I( \e^2 p_2),3\cdots,n \right)-M_n\left( \bar{\phi}_I( \e^2 p_1),\; \phi^I(\e^2 p_2),3\cdots,n \right)\bigg]\bigg|_{\e\rightarrow0}\,,\\
\nonumber (\mathcal{N}=6)\;&&\bigg[M_n\left( \phi^{IJ}( \e^2 p_1),\; \bar{\phi}_{IJ}( \e^2 p_2),3\cdots,n \right)-M_n\left( \bar{\phi}_{IJ}( \e^2 p_1),\; \phi^{IJ}(\e^2 p_2),3\cdots,n \right)\bigg]\bigg|_{\e\rightarrow0}\,,
\eqae
where the pairs $(\phi,\bar{\phi})$, $(\phi^I,\bar{\phi}_{I})$, and $(\phi^{IJ},\bar{\phi}_{IJ})$ indicate the SU($\mathcal{N}$) singlet combination of the $2$, $10$ and $30$ scalars in $\mathcal{N}=4,5$ and $6$ supergravity theories respectively. We will refer to such extraction of the double-soft limit as ``anti-symmetrised extraction". Note that due to the fact that we are considering non-maximal supergravity theories, the on-shell degrees of freedom are carried by two distinct mulitplets, each can be considered as a particular truncation of the maximal theory~\cite{LessSUSY}, 
\eq
\Phi^{\mathcal{N}}=\Phi^{\mathcal{N}=8}|_{\eta^8,\cdots,\eta^{\mathcal{N}{+}1}\rightarrow0}\,, \quad \overline{\Phi}^{\mathcal{N}}=\int d\eta^8\cdots d\eta^{\mathcal{N}{+}1}\Phi^{\mathcal{N}=8}\,,
\eqe
where $\Phi^{\mathcal{N}=8}$ is the unique superfield for the $\mathcal{N}=8$ theory. As a consequence, the U(1)-generator $R_a$ has a different constant for the two distinct multiplets
\eq\label{U1Def}
R_{a}=\sum_{I}\eta_a^{I}\frac{\partial}{\partial \eta_a^I} \quad (a\in \Phi^{\mathcal{N}})\,, \quad\quad \bar{R}_{a}=\sum_{I}\eta_a^{I}\frac{\partial}{\partial \eta_a^I}-\mathcal{N}\quad (a\in \overline{\Phi}^{\mathcal{N}})\,.
\eqe
One can verify that all tree amplitudes vanish under the above refined U(1) generator, i.e
\eq
\left(\sum_{a\in \Phi^{\mathcal{N}}} R_a + \sum_{b \in \overline{\Phi}^{\mathcal{N}}} \bar{R}_b \right) M_n=0\,.
\eqe

\subsection{Double-soft limits of  $4\leq \mathcal{N}<8$ Supergravity}
Let us now demonstrate the validity of eq.(\ref{Predict}) for  $4\leq \mathcal{N}<8$ Supergravity. We will use the on-shell recursion formula introduced by Britto, Cachazo, Feng and Witten~\cite{BCFW} to generate the tree amplitudes of $\mathcal{N}=8$ supergravity and perform SUSY reduction. To guarantee the presence of a U(1) on the right-hand side of $[T,T]\sim H$, we choose two scalars from the $\mathcal{N}=8$ theory that form a singlet, for example:
$$(\phi^{1234}, \phi^{5678})\,.$$
In terms of $\mathcal{N}=4, 5, 6$ representation, this would correspond to the scalar pairs $(\phi,\bar{\phi})$, $(\phi_5, \bar{\phi}^5)$ and $(\phi_{56}, \bar{\phi}^{56})$ respectively. The double-soft limits of scalar pairs that do not contain such singlet contribution can be derived similarly without the complication of soft-graviton divergence. We will simply present the final result for these cases.

We begin by considering the double-soft limit of the following two amplitudes: 
\beqa
\label{initial}
\begin{array}{cl}
(a)&~~\displaystyle\left(\int d^4\eta^{1234}_1 d^4\eta^{5678}_2 M^{\mathcal{N}=8}_n\right)\Bigg|_{p_1\rightarrow \e^2 p_1\atop{p_2\rightarrow \e^2 p_2}}\,.\\
(b)&~~\displaystyle \left(\int d^4\eta^{1234}_2 d^4\eta^{5678}_1M^{\mathcal{N}=8}_n\right)\Bigg|_{p_1\rightarrow \e^2 p_1\atop{p_2\rightarrow \e^2 p_2}}\,.
\end{array}
\eeqa
For both cases, the double-soft limits are divergent due to the presence of the soft-graviton pole. To extract the finite term we consider the difference
\eq
M_n\left(\phi^{1234}(1)\phi^{5678}(2)\cdots\right)-M_n\left(\phi^{5678}(1)\phi^{1234}(2)\cdots\right)\bigg|_{p_1,p_2\rightarrow \e^2 p_1, \e^2 p_2}\,.
\eqe
This anti-symmetrised extraction procedure will allow us to isolate the finger print of the U(1) duality group.

We begin with case $(a)$ in eq.(\ref{initial}). The BCFW shift is given by
$$|\hat{1}\rangle=|1\rangle+z|n\rangle\,, \;\; |\hat{n}]=|n]-z|1]\,,\;\; \eta_{\hat{n}}=\eta_n-z\eta_1 \, .$$
We will take the soft limit by setting $\lambda_{1,2}\rightarrow \epsilon \lambda_{1,2}$, $\tilde\lambda_{1,2}\rightarrow \epsilon \tilde\lambda_{1,2}$. Since the amplitudes vanish in the single-soft scalar limit, legs $1$ and $2$ must be on the same subamplitude of the BCFW diagram. However, since leg $1$ is shifted, plugging the explicit solution for $z$ in the generic multiplicity will render the momentum of leg $1$ hard. In this case, the subamplitude is again in a single-soft scalar limit, and thus vanishes. The only exception is when both legs $1$ and $2$ are on a three- or a four-point amplitude. For these diagrams, the propagators vanish in the limit, and one can potentially encounter $0/0$ cancellations. \\
~\\

\noindent $\bullet$ \textbf{BCFW diagram with a 4-point subamplitude}

We first begin with the latter and consider the following BCFW diagram:
$$\includegraphics[scale=0.65]{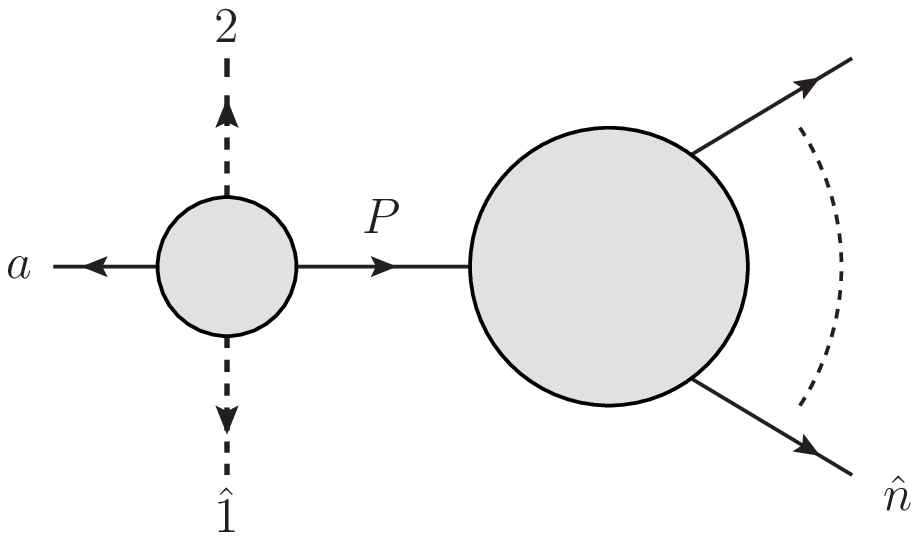}\,.$$
The contribution of this diagram is given as
\eqa\label{Initial}
\nonumber &&\int d^8\eta_{{P}} M_4(\hat{1},2,{P},a)\frac{1}{p_{1,2,a}^2}M_{n-2}(-{P},\cdots, \hat{n})\\
& &=\frac{s_{a2}\langle \hat{1}P\rangle^8}{\langle \hat{1}a\rangle\langle a2\rangle\langle 2P\rangle\langle P\hat{1}\rangle\langle \hat{1}2\rangle\langle 2a\rangle\langle aP\rangle\langle P\hat{1}\rangle p^2_{12a}} \int d^8\eta_{{P}} \delta^8_A \delta^8_B M_{n{-}2}({-}{P},\cdots, \hat{n})\,.
\eqae
The explicit solution to the shifted variable $z$ is given by
\eqa\label{zpsol}
\nonumber z_p &=&-\epsilon\frac{2p_a\cdot(p_1+p_2)}{\langle  n |a|1]}+\mathcal{O}(\epsilon^3)\,.
\eqae
Since $z_P$ is of order $\epsilon$, $|\hat{1}\rangle \sim \mathcal{O}(\epsilon)$, thus the deformed $p_{\hat{1}}$ is still soft in this channel. The spinors for the internal momentum is normalized as $ | P\rangle \sim \frac{p_a|1]}{[a1]}$ and $[  P\normalsize |  \sim -\frac{\langle n|p_a}{\langle na\rangle}$. The fermonic delta-functions are given as
\eq
\delta_A^8 :=\delta^8\left(\eta_{P} +\epsilon\frac{\langle \hat{1}2\rangle}{\langle \hat{1} {P}\rangle}\eta_2+\frac{\langle \hat{1}a\rangle}{\langle \hat{1} {P}\rangle}\eta_a \right),\;\;\quad \delta_B^8 :=\delta^8\left(\eta_1+\frac{\langle {P} 2\rangle}{\langle {P} \hat{1}\rangle}\eta_2+\epsilon\frac{\langle {P} a\rangle}{\langle {P} \hat{1}\rangle}\eta_a \right)\,.
\eqe
For convenience, we have explicitly written out the $\e$ dependence.

It is straightforward to see that in the double-soft limit the bosonic pre-factor in eq.(\ref{Initial}) is of order $\epsilon^{-2}$. We can use $\delta_A^8$ to localize the $d\eta_P$ integral and the net effect is $\eta_{{-}P}$ in $M_{n-2}$ is replaced by $(\epsilon\frac{\langle \hat{1}2\rangle}{\langle \hat{1}P\rangle}\eta_2+\eta_a)$. Thus the integrand in eq.(\ref{Initial}) can be written as
\eqa
\nonumber &&\hspace{-1cm}\frac{s_{a2}\langle \hat{1}P\rangle^8}{\langle \hat{1}a\rangle\langle a2\rangle\langle 2P\rangle\langle P\hat{1}\rangle\langle \hat{1}2\rangle\langle 2a\rangle\langle aP\rangle\langle P\hat{1}\rangle p^2_{12a}}
\\
&&~~~~~\times {\exp}\left(-\epsilon\frac{\langle \hat{1}2\rangle}{\langle \hat{1}P\rangle}\eta_2\frac{\partial}{\partial\eta_a}\right) {\exp} \left(- \e z_P\eta_1\frac{\partial}{\partial\eta_n}\right)
{\exp} \left(- \e^2 z_P\tilde{\lambda}_1\frac{\partial}{\partial \tilde{\lambda}_n}\right) M_{n{-}2}\,,
\eqae
where $M_{n{-2}}$ in the last line is now the unshifted $(n{-}2)$-point amplitude.

Now we want to pick the scalar components on legs $1$ and $2$, which entails computing 
\eq\label{Fruit0}
\int d^4\eta_2 d^4\eta_1 \;\delta^8_B\; {\exp} \left(-\epsilon\frac{\langle \hat{1}2\rangle}{\langle \hat{1}P\rangle}\eta_2\frac{\partial}{\partial\eta_a}\right) {\exp} \left( -\e z_P\eta_1\frac{\partial}{\partial\eta_n}\right)
{\exp} \left(- \e^2 z_P\tilde{\lambda}_1\frac{\partial}{\partial \tilde{\lambda}_n}\right)
 M_{n{-}2}\,.
\eqe
If all four $\eta_2$'s and four $\eta_1$'s came from $\delta^8_B$, we obtain the singlet contribution, which is divergent as $1/\epsilon^2$. It turns out that the leading divergent term as well as the subleading contribution are the same for both (a) and (b) in eq.(\ref{initial}), and thus cancel under the anti-symmetrized extraction. To get a non-vanishing result, one must pull down one factor of $\eta$ from the exponent. This will result in finite contributions, as it brings down a factor of $\epsilon$, along with the remaining  $\epsilon$ factor associated with $\eta_a$ in $\delta^8_B$. Thus for finite contribution we can either pull down an $\eta_2$ or an $\eta_1$ from the exponent.

Let's begin with taking an $\eta_2$ from the exponent. This means that $\delta^8_B$ contributes 4 $\eta_1$'s, 3 $\eta_2$'s and left with an $\eta_a$ unintegrated. What we then get is
\eqa
\nonumber&&\hspace{-0.8cm}-\frac{s_{a2}\langle \hat{1}P\rangle^8}{\langle \hat{1}a\rangle\langle a2\rangle\langle 2P\rangle\langle P\hat{1}\rangle\langle \hat{1}2\rangle\langle 2a\rangle\langle aP\rangle\langle P\hat{1}\rangle 2p_a\cdot(p_1+p_2)}\left(\frac{\langle P2\rangle}{\langle P\hat{1}\rangle}\right)^3\frac{\langle Pa\rangle}{\langle P\hat{1}\rangle}\frac{\langle \hat{1}2\rangle}{\langle \hat{1}P\rangle}\sum_{I=5}^8\eta^I_a\frac{\partial}{\partial\eta^I_a}M_{n-2}\\
&&\hspace{-0.8cm}=\frac{p_a\cdot p_2}{ p_a \cdot (p_1+p_2)}\sum_{I=5}^8\eta^I_a\frac{\partial}{\partial\eta^I_a}M_{n-2}\,.
\eqae
Next, we consider bringing down an $\eta_1$ instead. In this case, $\delta^8_B$ contributes 3 $\eta_1$'s and 4 $\eta_2$'s, and leaves an $\eta_a$ unintegrated. Using the explicit form of $z_p$ in eq.(\ref{zpsol}) we have
\eqa
\nonumber&&-\frac{s_{a2}\langle \hat{1}P\rangle^8}{\langle \hat{1}a\rangle\langle a2\rangle\langle 2P\rangle\langle P\hat{1}\rangle\langle \hat{1}2\rangle\langle 2a\rangle\langle aP\rangle\langle P\hat{1}\rangle }\frac{1}{\langle n|p_a|1] }\left(\frac{\langle P2\rangle}{\langle P\hat{1}\rangle}\right)^4\frac{\langle Pa\rangle}{\langle P\hat{1}\rangle}\sum_{I=1}^4\eta^I_a\frac{\partial}{\partial\eta^I_n}M_{n-2}\\
&&=\frac{[a2]}{\langle \hat{1}2\rangle }\frac{\langle a2\rangle^2}{\langle n|p_a|1] }\sum_{I=1}^4\eta^I_a \frac{\partial}{\partial\eta^I_n}M_{n-2}\,.
\eqae
Using the explicit representation for $|\hat{1}\rangle$, the above can be written as
\eq\label{Fruit1}
\frac{[a2]}{\langle \hat{1}2\rangle }\frac{\langle a2\rangle^2}{\langle n|p_a|1] }\sum_{I=1}^4\eta^I_a \frac{\partial}{\partial\eta^I_n}M_{n-2}=\frac{[a2]\langle 2a\rangle}{\langle n|2+1|a]}\sum_{I=1}^4\eta^I_a\frac{\partial}{\partial\eta^I_n}M_{n-2}\,.
\eqe

\noindent $\bullet$ \textbf{BCFW diagram with a 3-point subamplitude}

\begin{figure}[t]
\begin{center}
\includegraphics[scale=0.65]{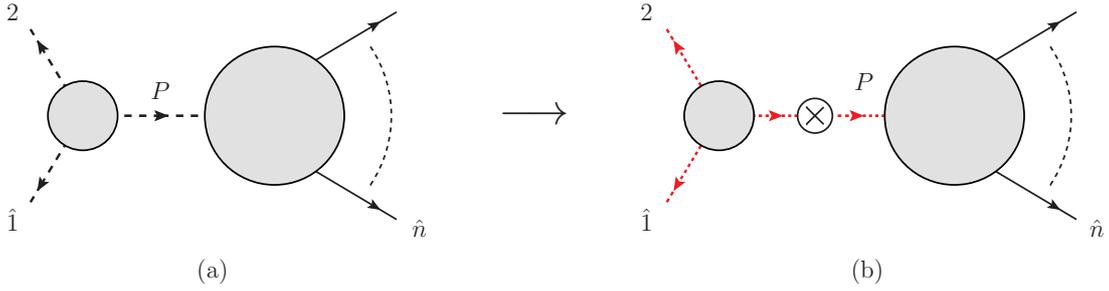}
\caption{ (a) The BCFW diagram where the two soft legs are attached to a three-point sub amplitude. (b) Due to the soft kinematics, the diagram factorizes into a three-point times an $n{-}1$-point amplitude in the single soft limit.}
\label{OnlyFig}
\end{center}
\end{figure}

Let us now consider the following BCFW diagram with a 3-point subamplitude. The relevant diagram is displayed in diagram (a) of fig.\ref{OnlyFig}, which yields 
\eqa\label{3ptInt}
&{}&M_3(\hat{1},2,P)\frac{1}{p_{12}}M_{n{-}1}(-P,\dots,\hat{n})\\
&=& 
\frac{\delta^8_A([12]\eta_P+[2P]\eta_1+[P1]\eta_2)}{[12]^2[2P]^2[P1]^2s_{12}}\exp\left(-\epsilon z_P\eta_1\frac{\partial}{\partial \eta_n}\right)
{\exp} \left(- \e^2 z_P\tilde{\lambda}_1\frac{\partial}{\partial \tilde{\lambda}_n}\right)
 M_{n{-}1}(P,\dots,n)\,. \nonumber
\eqae
The internal momentum and the solution for $z_P$ is given as
\eq\label{zsol}
|P\rangle=\e|2\rangle\,,\quad[P|=-\e\left([2|+\frac{\langle n1\rangle}{\langle n2\rangle}[1|\right),\quad z_P=-\e\frac{\langle 12\rangle}{\langle n2\rangle}\,.
\eqe
Again, we have explicitly written out the $\e$ dependence for the exponents. A new feature is that the $(n{-}1)$-point amplitude on the RHS is in fact divergent due to the presence of a soft graviton. In particular since $P$ is soft, the RHS of the diagram is an $(n{-}1)$-pt amplitude in the single-soft-limit as illustrated in diagram (b) of fig.(\ref{OnlyFig}),. The soft-graviton divergence of the $M_{m+1}$ is given by~\cite{Weinberg, CS}:
\eq\label{SoftExpand}
\mathcal{M}_{m+1}(1,\cdots,m,\e^2 s)=\frac{1}{\epsilon^2}\mathcal{S}_G^{(0)} \mathcal{M}_{m}(1,\cdots,n)+\mathcal{S}_G^{(1)}\mathcal{M}_{m}(1,\cdots,n)\,.
\eqe
As the three-point amplitude behaves as $\e^4$ while the propagator as $\e^{-2}$, the only term we need to consider is $\mathcal{S}^{(0)}_G$. The explicit supersymmetric single-soft operator of an $(n{-}1)$-point supersymmetric amplitude is known to be~\cite{HHW}
\eqa
\nonumber M_{n{-}1}(P,3,\dots,n)~&\rightarrow&~{1 \over \epsilon^2 } \sum_{a=3}^{n}\frac{\langle aP \rangle [na]^2}{[nP]^2[aP]}\de_B^8\left(\eta_P+\epsilon\frac{[nP]}{[an]}\eta_a+\epsilon\frac{[Pa]}{[an]}\eta_n\right)M_{n-2}(3,\dots,n)\,,
\eqae 
where we only keep the relevant leading term.

Similar to the previous analysis, in order for there to be a non-vanishing result after the anti-symmetrised extraction, we must take one of the $\eta$'s from the exponents. In eq.(\ref{3ptInt}) we can only choose $\eta_1$ and thus $\delta^8_A$ contributes 3 $\eta_1$'s and $4$ $\eta_2$'s. Again using $\delta^8_B$ to localise $\eta_P$, the result is:   
\eqa
\frac{[12][2P]^3[P1]^4}{[12]^2[2P]^2[P1]^2}\frac{z_P}{s_{12}}\sum_{a=3}^{n-1}\frac{\langle aP \rangle [na]^2}{[nP]^2[Pa]}\sum_{I=1}^4\left(\frac{[nP]}{[an]}\eta^I_a+\frac{[Pa]}{[an]}\eta^I_n\right)\frac{\partial}{\partial \eta^I_n} M_{n-2}(3,\dots,n)\,.
\eqae
Using eq.(\ref{zsol}) the term above can be rewritten as, to leading order,
\eqa
\hspace{-0.8cm}-\sum_{I=1}^4\sum_{a=3}^{n-1}\left[\frac{\langle 1n\rangle \langle 2a\rangle [na][12]}{\langle n|(p_1+p_2)|a]p_n\cdot(p_1+p_2) }\eta^I_a-\frac{\langle 1 n\rangle[na][12]\langle2 a\rangle}{ [p_n\cdot(p_1+p_2) ]^2}\eta^I_n\right]\frac{\partial}{\partial \eta^I_n} M_{n-2}(3,\dots,n)\,.
\eqae
Note that the second term in the soft limit vanishes due to the momentum conservation, and the first term can be combined with the previous BCFW result in eq.(\ref{Fruit1}) as
\eqa
&&\sum_{a=3}^{n-1}\left[{\langle 2 a\rangle [a2]  \over \langle n|(p_1+p_2)|a]}-\frac{\langle 1n\rangle \langle 2a\rangle [na][12]}{\langle n|(p_1+p_2)|a]p_n\cdot(p_1+p_2) }\right]\sum_{I=1}^4\eta^I_a\frac{\partial}{\partial \eta^I_n} M_{n-2}(3,\dots,n) \cr
&&=\sum_{a=3}^{n{-}1}\frac{\langle 2 a\rangle [a2]p_n\cdot(p_1+p_2)-\langle 1n\rangle \langle 2a\rangle [na][12]}{\langle n|(p_1+p_2)|a] p_n\cdot(p_1+p_2)}\sum_{I=1}^4
\eta^I_a\frac{\partial}{\partial \eta^I_n} M_{n-2}(3,\dots,n)\,.
\eqae
Using the $n{-}2$-point super momentum conservation, the numerator can be simplified such that one has
\eqa
\nonumber\hspace{-0.8cm}\sum_{I=1}^4 \sum_{a=3}^{n-1}\frac{[n2]\langle 2 a\rangle}{(p_1+p_2)\cdot p_n}\eta^I_a\frac{\partial}{\partial \eta^I_n} M_{n-2}(3,\dots,n)=-\sum_{I=1}^4\frac{p_n\cdot p_2}{(p_1+p_2)\cdot p_n}\eta^I_n\frac{\partial}{\partial \eta^I_n} M_{n-2}(3,\dots,n)\,,\\
\eqae
Put everything together, we find that the difference for scenario (a) and (b) in eq.(\ref{initial}) is given by
\eqa
\nonumber&&\hspace{-1.5cm}M_n\left(\phi^{1234}(1)\phi^{5678}(2)\cdots\right)-M_n\left(\phi^{5678}(1)\phi^{1234}(2)\cdots\right)\bigg|_{p_1,p_2\rightarrow \e^2 p_1, \e^2 p_2}\\
\nonumber&&\hspace{-1.2cm}=\bigg[\left(\sum_{a=3}^{n{-}1}\frac{p_a\cdot p_2}{ p_a \cdot (p_1+p_2)}\sum_{I=5}^8\eta^I_a\frac{\partial}{\partial\eta^I_a}-\frac{p_n\cdot p_2}{p_n \cdot (p_1+p_2)}\sum_{I=1}^4\eta^I_n\frac{\partial}{\partial \eta^I_n} \right)\\
&&~~~-\left(\sum_{a=3}^{n{-}1}\frac{p_a\cdot p_2}{ p_a \cdot (p_1+p_2)}\sum_{I=1}^4\eta^I_a\frac{\partial}{\partial\eta^I_a}-\frac{p_n\cdot p_2}{p_n \cdot (p_1+p_2)}\sum_{I=5}^8\eta^I_n\frac{\partial}{\partial \eta^I_n} \right)\bigg]M_{n-2} \cr
&&\hspace{-1.2cm}=
\bigg[\sum_{a=3}^{n}\frac{p_a\cdot p_2}{ p_a \cdot (p_1+p_2)}
\left( \sum_{I=5}^8\eta^I_a\frac{\partial}{\partial\eta^I_a}- \sum_{I=1}^4\eta^I_a\frac{\partial}{\partial\eta^I_a}  \right)\bigg]M_{n-2}\,.
\eqae
Thus we see that after anti-symmetrized extraction, the double-soft limit results in single-site U(1)-generators acting on a lower-point amplitude.

We now perform the SUSY reduction to $\mathcal{N}<8$. In the reduction, for each leg one needs to choose between integrating away $d\eta^{\mathcal{N}{+}1}\cdots d\eta^8$ to obtain the $\overline{\Phi}$ multiplet, or setting all $\eta^{\mathcal{N}{+}1}\cdots \eta^8$s to be zero for the $\Phi$ multiplet. Denote the $(n{-}2)$ points in two sets, with $\alpha\in\overline{\Phi}$ and $\beta\in\Phi$. For the legs in $\alpha$, integrating $d\eta^{\mathcal{N}{+}1}\cdots d\eta^8$ will leave behind:
\eqa\label{Prelim1}
\hspace{-0.8cm} (\alpha):&&-\frac{p_a\cdot p_2}{ p_a\cdot(p_1+p_2)}\left(\sum_{I=1}^4\eta^I_a\frac{\partial}{\partial\eta^I_a}-\sum_{J=5}^{\mathcal{N}}\eta^J_a\frac{\partial}{\partial\eta^J_a}-8+\mathcal{N}\right)\int d\eta^{\mathcal{N}{+}1}\cdots d\eta^8M_{n-2}\,,
\eqae
where we've used the identity $\int d\eta\, \eta\frac{\partial}{\partial \eta} \,\ast=\int d\eta \,\ast$. On the other hand for the legs in $\beta$, we have:
\eqa\label{Prelim2}
\nonumber&&\hspace*{-1.37cm}(\beta):~\left[-\frac{p_a\cdot p_2}{ p_a\cdot(p_1+p_2)}\left(\sum_{I=1}^4\eta^I_a\frac{\partial}{\partial\eta^I_a}-\sum_{I=5}^{\mathcal{N}}\eta^I_a\frac{\partial}{\partial\eta^I_a}\right)M_{n-2}\right]\bigg|_{\eta^{\mathcal{N}+1}\cdots \eta^8\rightarrow0}\\
&&~~~~~~~~=-\frac{p_a\cdot p_2}{ p_a\cdot(p_1+p_2)}\left(\sum_{I=1}^4\eta^I_a\frac{\partial}{\partial\eta^I_a}-\sum_{I=5}^{\mathcal{N}}\eta^I_a\frac{\partial}{\partial\eta^I_a}\right)\left[M_{n-2}\right]|_{\eta^{\mathcal{N}+1}\cdots \eta^8\rightarrow0}\,.
\eqae
The above result is precisely eq.(\ref{Predict}). To see this, recall that 
\eq
\left(R_a\right)^{I}\,_I=\eta_a^I\frac{\partial}{\partial \eta_a^I}-\frac{\delta^I_I}{\mathcal{N}}\left(\sum_{J=1}^{\mathcal{N}}\eta_a^J\frac{\partial}{\partial \eta_a^J}\right)\,,
\eqe
where the repeated indices are not summed over. Combined with the definition of the U(1) generators in eq.(\ref{U1Def}), we can see that eq.(\ref{Prelim1},\,\ref{Prelim2}) are simply:
\eqa\label{Predict2}
\nonumber (\mathcal{N}=4)\;&&\bigg[M_n\left( \phi( \e^2 p_1),\; \bar{\phi}( \e^2 p_2),3\cdots,n \right)-M_n\left( \bar{\phi}( \e^2 p_1),\; \phi(\e^2 p_2),3\cdots,n \right)\bigg]\bigg|_{\e\rightarrow0}\\
&&=\sum_{a=3}^{n} \frac{p_a\cdot (p_1-p_2)}{2p_a\cdot(p_1+p_2)}\left(R_{a}\right) M_{n-2}+\mathcal{O}(\e)\,,\\
\nonumber (\mathcal{N}=5)\;&&\bigg[M_n\left( \phi^I( \e^2 p_1),\; \bar{\phi}_I( \e^2 p_2),3\cdots,n \right)-M_n\left( \bar{\phi}_I( \e^2 p_1),\; \phi^I(\e^2 p_2),3\cdots,n \right)\bigg]\bigg|_{\e\rightarrow0}\\
\nonumber&&=\sum_{a=3}^{n} \frac{p_a\cdot (p_1-p_2)}{2p_a\cdot(p_1+p_2)}\left(\left(R_a\right)^I_I+\frac{\mathcal{N}-8}{2\mathcal{N}}\delta^{I}_{I}R_{a}\right) M_{n-2}+\mathcal{O}(\e)\,,\\
\nonumber (\mathcal{N}=6)\;&&\bigg[M_n\left( \phi^{IJ}( \e^2 p_1),\; \bar{\phi}_{IJ}( \e^2 p_2),3\cdots,n \right)-M_n\left( \bar{\phi}_{IJ}( \e^2 p_1),\; \phi^{IJ}(\e^2 p_2),3\cdots,n \right)\bigg]\bigg|_{\e\rightarrow0}\\
\nonumber&&=\sum_{a=3}^{n} \frac{p_a\cdot (p_1-p_2)}{2p_a\cdot(p_1+p_2)}\left(\left(R_a\right)^I_I\delta^{J}_{J}+\left(R_a\right)^J_J\delta^{I}_{I}+\frac{\mathcal{N}-8}{2\mathcal{N}}\delta^{IJ}_{IJ}R_{a}\right) M_{n-2}+\mathcal{O}(\e)\,,
\eqae
where $M_{n-2}$s in each line correspond to the amplitudes in $\mathcal{N}=4,5,6$ supergravity theories respectively. This completes our proof.

\section{Soft scalars in three-dimensional supergravity } \label{section:3D}

\subsection{Review of $\mathcal{N}=16$ supergravity }

In three dimensions, the graviton does not have physical degrees of freedom. If one dimensionally reduces four-dimensional Einstein-Hilbert gravity, the two physical degrees of freedom become scalars in three dimensions. Thus the gravity amplitudes under consideration correspond to the scattering of scalars, and their supersymmetric partners, with their interactions mediated by gravitons. Thus in a sense the system is very much like Chern-Simons matter theories, where the physical matter fields interact through a topological gauge field, and both systems can be considered as a perturbation of a topological theory. Like their higher dimensional parents, the scalars in the supersymmetric theories are coordinates of a coset manifold, and the purely scalar part of the action is given by a non-linear sigma model, i.e. three-dimensional supergravity theories are really local supersymmetric non-linear sigma models. Unlike their higher-dimensional counter part, in three-dimensions, \textit{all} bosonic degrees of freedom are governed by this coset structure.

The theory we will discuss here is the $\mathcal{N}=16$ theory constructed by Marcus and Schwarz~\cite{E8}. 128 scalars and 128 fermions in the theory transform under inequivalent spinor representation of SO(16) R-symmetry. Due to the fact that the physical degrees of freedom are in the spinor representation, they have to come in pairs to form a singlet, and hence only even-multiplicity S-matrix is non-trivial. The scalars parametrize the coset space E$_{8(8)}$/SO(16), where the extra $8$ in E$_8$ denotes that the difference between non-compact and compact generators is 8.  

The on-shell degrees of freedom are encoded in a superfield that is a function of $\eta^I$ where $I=1$, $2$, $\dots$, $8$ transforms as the fundamental representation of U(8). In this language, the rest $56$ of $120$ SO(16) generators are non-linearly realized:
\eq\label{Gen1}
 (\mathbf{28}):~\sum_{a} R_a^{IJ} \equiv  \sum_{a}\eta_a^I\eta_a^J\,,~~~~ (\mathbf{28}):~  \sum_a (R_a)_{IJ} \equiv  \sum_{a}\frac{\partial}{\partial\eta_a^I}\frac{\partial}{\partial\eta_a^J}\,,
\eqe 
where the numbers in the parenthesis denotes the numbers of generators. For completeness, the linearly realized  U(8)=SU(8)$\times$U(1) is given as:
\eq\label{Gen2}
\begin{array}{cl}
(\mathbf{63}):&~\displaystyle\sum_a (R_a)^I~_J \equiv  \sum_a \left[\eta^I_a\frac{\partial}{\partial\eta_a^J}-\frac{\delta^{I}_J}{8}\left(\sum_{K}\eta_a^{K}\frac{\partial}{\partial \eta_a^K}\right)\right]\,,\\
 \mathbf{(1)}:&~\displaystyle\sum_a R_a \equiv  \sum_{a} \left(\sum_{K}\eta_a^{K}\frac{\partial}{\partial \eta_a^K}-4\right)\,.
\end{array}
\eqe
The 128$\oplus$128 states are grouped in a SU(8) singlet scalar super field:
\beqa
\label{Pxi}
\Phi&=&\xi+\frac{1}{2}\xi_{IJ}(\eta^2)^{IJ}+\frac{1}{4!}\xi_{IJKL}(\eta^4)^{IJKL}+\frac{1}{6!}\xi_{IJKLMN}(\eta^6)^{IJKLMN}+\bar{\xi}\eta^8\,,
\eeqa
where we have only denoted the scalar states. Since all scalars participate in the duality group, we anticipate a much richer double-soft structure. In particular we expect:
\beqa
\label{D=3Predict}
\begin{array}{rcl}
 [\xi_{I_{1}I_{2}}\;,\;\bar{\xi}]_{s}M_{n}&\sim & -2R_{I_1I_2}M_{n-2}\,,\\
\left[\xi_{I_{1}I_{2}I_{3}I_{4}I_{5}I_{6}}\;,\;\xi_{J_1J_2J_3J_4}\right]_sM_{n}&\sim &6\epsilon_{I_{1}I_{2}I_{3}I_{4}I_{5}I_{6}[J_1J_2}R_{J_3J_4]}M_{n-2}\,,\\
\left[\xi_{I_{1}I_{2}}\;,\; \xi_{J_1J_2J_3J_4}\right]_sM_{n}&\sim & \e_{I_{1}I_{2}J_1J_2J_3J_4K_1K_2}R^{K_1K_2}M_{n-2}\,,\\
\left[\xi_{I_{1}I_{2}I_{3}I_{4}I_{5}I_{6}}\;, \;\xi\right]_sM_{n}&\sim &\e_{I_{1}I_{2}I_3I_4I_5I_6K_1K_2}R^{K_1K_2}M_{n-2}\,,\\
\left[\xi_{I_{1}I_{2}I_{3}I_{4}I_{5}I_{6}}\;, \;\xi_{J_{1}J_{2}}\right]_sM_{n}&\sim &\left(4\e_{I_{1}I_{2}I_{3}I_{4}I_{5}I_{6}J[J_{1}} R^{J}\,_{J_2]}+\frac{1}{2}\e_{I_{1}I_{2}I_{3}I_{4}I_{5}I_{6}J_{1}J_2}R \right)M_{n-2}\,,\\
  \left[\xi_{I_{1}I_{2}I_{3}I_{4}}, \;\xi_{J_{1}J_{2}J_3J_4}\right]_sM_{n}&\sim &\left(8\e_{I_{1}I_{2}I_{3}I_{4}J[J_{1}J_{2}J_3} R^{J}\,_{J_4]} \right)M_{n-2}\,,\\
\left[\xi\;,\; \bar{\xi}\right]_s 
M_{n}&\sim & RM_{n-2}\,, 
\end{array}
\eeqa
where $[\; , \;]_s$ indicates we are taking the antisymmetrised double-soft limit on the amplitudes with respect to the two scalars indicated in the brackets. As we will show in the following, the double-soft-limit of the tree and one-loop amplitudes indeed behave in the above fashion with the proportionality factor given by $p_a\cdot(p_1-p_2)/2p_a\cdot (p_1+p_2)$\,.

In the following, we will utilise the BCFW recursion, which has been applied to study scattering amplitudes in three-dimensional ABJM theory~\cite{3DRecur}. The applicability of BCFW for the $\mathcal{N}=16$ supergravity can be argued by comparing with its four-dimensional parent. Note that in the large-$z$ limit of three-dimensional BCFW correspond to boosting the momenta of the shifted legs along a null direction, with a proportionality given by $z^2$, exactly the same as its four-dimensional parent. Since three-dimensional kinematics is simply a special limit, the large-$z$ asymptotics can be deduced from four dimensions and one concludes that maximal supergravity behaves as $z^{-4}$ asymptotically.\footnote{This is also consistent with the observation in~\cite{David} that permutation invariance implies that at large-$z$, the amplitude behaves as $z^{4k}$. Since for the maximal theory all degrees of freedom are in the same multiplet, permutation invariance is satisfied.}
 \subsection{The double-soft limit: Tree level}

As we have seen from the previous section, the vacuum structure of supergravity theories in four dimensions can be explored by analysing the scattering amplitudes in the double-soft-scalar limit. As a reflection of the coset space structure, 
an $n$-point amplitude in the limit reduces to a rotation operator of R-symmetry acting on an $(n{-}2)$-point scattering amplitude. In this subsection, we will perform a similar analysis for $\mathcal N=16$ supergravity in three dimensions. As we mentioned earlier, due to the presence of a U(1) in the isotropy group, we will again have divergent contributions from singlet, which requires the antisymmetric extraction procedure introduced previously.

Again BCFW recursion relations, now in three dimensions, are the main tool for our analysis. We denote the BCFW shifts in 3D as~\cite{3DRecur},
\beqa
\label{bcfw3d}
\begin{array}{rclrcl}
\hat \lambda_1&=& c\lambda_1+s\lambda_n\,,&~~\hat \lambda_n&=&-s\lambda_1+c\lambda_n\,,\\
\hat \eta_1&=& c\eta_1+s\eta_n\,,&~~\hat \eta_n&=&-s\eta_1+c\eta_n \, ,
\end{array}
\eeqa
where $c^2+s^2=1$ required by the momentum conservation. It is convenient to solve the orthogonal constraint by introducing a parameter $z$, with
\beq
c= {1 \over 2} \left( z + 1/z \right) \, , \quad s= {i \over 2} \left( z - 1/z \right) \, ,
\eeq
or $z = c- i s$. Moreover, $c(z)$ and $s(z)$ are fully determined by the on-shell condition for internal momentum $P$ in a BCFW diagram. There are four sets of solutions, denoted as $z_j$ with $j=1,2,3,4$, but only two of them are linearly independent, the other two are related by an overall sign. As a consequence of the fact that there is a single bosonic superfield,\footnote{In three dimensions the little group is $Z_2$, and it acts on the on-shell variable as $\lambda\rightarrow -\lambda$. Thus there are only two types of particles, bosons which is a $Z_2$ singlet and fermions which are $Z_2$ odd. The $\mathcal{N}=16$ superfield is a bosonic superfield.} which implies the factorization of amplitudes $A_m(z) A_n(z)=A_m(-z) A_n(-z)$, and the linear dependence of the four set of solutions, it allows us to express the BCFW representation of amplitudes purely in terms of two out of four sets of solutions:
\beqa
A_n=\sum_f \int d^{\mathcal N}\eta_I A_L(z_{1,f};\eta_I)\frac{H(z_{1,f},z_{2,f})}{P_{12\dots i}^2}A_R(z_{1,f};i\eta_I)+(z_{1,f}\leftrightarrow z_{2,f})\,,
\eeqa
where
\beqa\label{HDef}
H(x,y)\equiv \frac{x^2(y^2-1)}{x^2-y^2} \, .
\eeqa

Before starting our investigation on the double-soft limit, we like to show that the amplitudes vanish in the single-soft limit, as the consequence of soft ``pion" theorem. This fact can be seen most easily by BCFW recursion relations. First of all, the totally permutation symmetric four-point amplitude is given as~\cite{Song}, 
\beqa
M_4(1,2,3,4)=\frac{\de^{16}( \sum_i \lambda_i \eta_i )\de^3(\sum_i p_i)}{\braket{1}{2}^2\braket{2}{3}^2\braket{3}{1}^2}\,.
\eeqa 
Expanding $\de^{16}( \sum_i \lambda_i \eta_i )$ out, it is straightforward to see that the amplitude vanishes as $\epsilon^2$ at the single-soft limit, say $\lambda_1 \rightarrow  \epsilon \lambda_1$. For a general higher-point amplitude, we can represent the amplitude by BCFW recursion relations with two shifted legs not involving the soft particle. Recursively apply the recursion relation, one can always reduce the amplitude into four-point ones, which we have just proved behaving as $\epsilon^2$, while the propagator in the BCFW diagram is always finite. Thus we conclude that any amplitudes in $\mathcal{N}=16$ supergravity vanish in the single-soft limit as $\epsilon^2$ in the limit. We will comment on the single-soft limit with more details in the next section.

Let us now consider the double-soft limit. Because the amplitudes in the single-soft-scalar limit vanish, and only even-point amplitudes are allowed in the theory, the only relevant BCFW diagram is a four-point amplitude with two soft legs glued with an $(n{-}2)$-point amplitude
\beqa
\notag   M_{n}\Big|_{\lambda_1\rightarrow 0\atop{\lambda_2\rightarrow 0}}&=&\sum_{a=3}^{n-1}\int d^{8}\eta_PM_4(\hat 1,a,2,P;z_1)\frac{H(z_1,z_2)}{(p_{1}+p_2+p_a)^2}M_{n-2}(-P,\dots,\hat n;z_1)+(z_1\leftrightarrow z_2)\,,\\
\label{n2}\eeqa
where two soft legs are chosen to be $1$ and $2$. Substituting the four-point amplitude into eq.\eqref{n2}, one gets
\beqa
 \label{Mn3d}
 M_{n}\Big|_{\lambda_1\rightarrow 0\atop{\lambda_2\rightarrow 0}}= \sum_{a=3}^{n-1}\mathcal H(z_1,z_2)\int d^{8}\eta_P\delta^{8}_{\rho}(z_1)\delta^{8}_{\sigma}(z_1)M_{n-2}(-P,\cdots, \hat{n};z_1)+(z_1\leftrightarrow z_2)\,,
\eeqa
where 
\beqa
\notag \delta_{\rho}^{8}\equiv\delta^{8}\left(\eta_P+\frac{\langle \hat{1}2\rangle}{\langle \hat{1}P\rangle}\eta_2+\frac{\langle \hat{1}a\rangle}{\langle \hat{1}P\rangle}\eta_a \right),\;\;\quad \delta_{\sigma}^{8}\equiv\delta^{8}\left(\hat\eta_1+\frac{\langle P2\rangle}{\langle P\hat{1}\rangle}\eta_2+\frac{\langle Pa\rangle}{\langle P\hat{1}\rangle}\eta_a\right)\,,
\eeqa
\beqa
\mathcal H(z_1,z_2)\equiv \frac{\langle \hat{1}P\rangle^{8}}{\langle \hat{1}a\rangle^2\langle a2\rangle^2\langle 2\hat 1\rangle^2}\frac{H(z_1, z_2)}{(p_1+p_2+p_a)^2} \,.
\eeqa
Note that four-point momentum conservation, with all legs on-shell, implies the following relations:
\beqa
\label{4pt}\braket{P}{\hat 1}=\pm \braket{a}{2}\,,~~\braket{P}{2}=\pm \braket{\hat 1}{a}\,,~~\braket{P}{a}=\pm \braket{2}{\hat 1}\,,
\eeqa
where $\pm$ signs correspond to the two on-shell solutions of internal momentum, $z_1$ and $z_2$, respectively. In the double-soft limit, we parametrize the spinors $\lambda_1\rightarrow \epsilon\lambda_1$ and $\lambda_2\rightarrow \epsilon\lambda_2$ by the parameter $\epsilon$. Only terms up to $\mathcal O(\epsilon^0)$ are important to us. To get precise contribution up to $\mathcal O(\epsilon^0)$, due to the leading  $\epsilon^{-2}$ contribution from the propagator, we expand remaining factors in the amplitudes up to $\mathcal{O}(\epsilon^2)$.

With eqs.\eqref{4pt}, we can solve $c_j$ and $s_j$ (for $j=1,2$) to the order of our interest, 
\beqa
\notag c_j&=& 1-\frac{\alpha_j^2}{2}\epsilon^2+\mathcal{O}(\epsilon^4)\,,\\
\label{cs} s_j&=& -\alpha_j\epsilon+\left[(\alpha_j+\alpha_j^*)\frac{\alpha_j^2}{4}- (\alpha_j-\alpha_j^*)\frac{\beta_j^2}{4}\right]\epsilon^3
+\mathcal{O}(\epsilon^5)\,,
\eeqa
where $\alpha_j$ and $\beta_j$  are defined as
\beqas
\alpha_1\equiv \frac{\langle 1a\rangle+ i \langle 2a\rangle}{\langle na\rangle},~~~~\beta_1\equiv \frac{\braket{1}{n}+ i\braket{2}{n}}{\braket{n}{a}},~~~~
\alpha_2\equiv\alpha_1^*,~~~~\beta_2\equiv\beta_1^* \, .
\eeqas 
Substituting the above solutions in relevant terms in eq.(\ref{Mn3d}), we find
\beqa
\de^8_\rho(z_j)&=&\de^{(8)}\left(\eta_P-i\left(\eta_a +(-1)^{j+1} i \beta_j\eta_2\epsilon+\frac{\beta^2_j\epsilon^2}{2}\eta_a\right)+\mathcal{O}(\epsilon^3)\right)\,,\\
\notag\de^8_\sigma(z_j)&=&\de^{(8)}\left(\left(1-\frac{\alpha_j^2}{2}\epsilon^2 \right)\eta_1+(-1)^{j+1}i\left(1+\frac{\beta_j^2}{2}\epsilon^2\right)\eta_2+\beta_j \eta_a\epsilon-\alpha_j\eta_n\epsilon+\mathcal{O}(\epsilon^3)\right)\,,\\
\notag 
\mathcal H_j&=&f_0(z_j)\epsilon^{-2}
+f_1(z_j)\epsilon^{-1}
+f_2(z_j)\epsilon^0
+\mathcal{O}(\epsilon)\,,~~~~~~~
f_0(z_j)\equiv-\frac{1}{\epsilon^2}\frac{\alpha_j-\alpha^*_j}{4\alpha_j\beta^2_j}\,,\eeqa
where $\mathcal H_i \equiv \mathcal H(z_i,z_j)$. Explicit forms of $f_1$ and $f_2$ are actually irrelevant under antisymmetric extraction. This will be clear shortly.
Carrying out the integration of $\eta_P$ on $\de_\rho$ in eq.\eqref{Mn3d}, the information of $\de_\sigma$ can be recast into an operator acting on the remaining $(n{-}2)$-point amplitudes, similarly for the BCFW shifted variables $\hat{\lambda}_{1,n}$ and $\hat{\eta}_{1,n}$. By doing so, we find
\beqa
\label{complM}  M_{n}\Big|_{\lambda_1\rightarrow \epsilon\lambda_1\atop{\lambda_2\rightarrow \epsilon \lambda_2}}&=&\sum_{a=3}^{n-1}\mathcal H(z_1,z_2)\de_\sigma^8(z_1)\exp\big[\mathcal U(z_1)\big] M_{n-2}(a,\dots, n;z_1)+(z_1\leftrightarrow z_2)\,,\\
\notag\mathcal U(z_j)&\equiv&\underbrace{ 
(-1)^{j+1} i\epsilon\beta_j \eta_2
\frac{\partial}{\partial \eta_a}+\epsilon^2 O_{\eta_a}(z_i)}_{\mathrm{integration\,of}\,\eta_P }+\underbrace{
\epsilon \alpha_j \eta_1
\frac{\partial}{\partial \eta_n}+\epsilon^2 O_{\eta_n}(z_i) + \epsilon^2 O_{\lambda_{1,n}}(z_i)}_{\mathrm{BCFW\,shifted}\,\hat{\lambda}_{1,n},\,\hat{\eta}_{1,n}}+\mathcal O(\epsilon^3)\,.
\eeqa
Here, $O_{\eta_a,\eta_n,\lambda_{1,n}}$ are differential operators dependent on $\partial_{\eta_a,\eta_n,\lambda_{1,n}}$\,.

To verify eq.(\ref{D=3Predict}), we will integrate away $m_1$ number of $\eta_1$'s and $m_2$ number of $\eta_2$'s, with the following possible choices: $(m_1,m_2)=(8,2)$, $(6,4)$, $(8,0)$, $(6,2)$, $(4,4)$, $(6,0)$, $(4,2)$, as well as those where $m_1\leftrightarrow m_2$. To simplify the notations, in the following, the amplitudes with soft scalars $\xi^{(m_A)}$ and $\xi^{(m_B)}$, where $\xi$s are defined in eq.\eqref{Pxi}, will be abbreviated as
\beqa
\int d^{8}\eta_2 d^{8}\eta_1 \eta_2^{8-m_B}\eta_1^{8-m_A} M_{n}(\xi^{(m_A)}(p_1),\xi^{(m_B)}(p_2),\dots)\Big|_{\lambda_1\rightarrow \epsilon\lambda_1\atop{\lambda_2\rightarrow \epsilon \lambda_2}}\equiv \mathcal M_n^{(m_A,m_B)},
\eeqa
 where $m_A$ and $m_B$ are the number of U$(8)$ indices carried by the soft particles with momentum $p_1$ and $p_2$, respectively, and the subscripts $A$ and $B$ of which are the set of U$(8)$ indices $\{A_i\}$ and $\{B_i\}$ labelled on the fields $\xi$'s. To see why we are interested in the aforementioned sets of $(m_A, m_B)$, a closer inspection of eq.(\ref{D=3Predict}) tells us that depending on the sum $m_A+m_B$, different SO(16) generators are expected on the RHS. In particular, we have:
\eqa
(m_A+m_B)=10:\; R_{IJ}\,, \quad(m_A+m_B)=8:\;R^{I}\,_{J}\,,\, R, \quad(m_A+m_B)=6:\; R^{IJ}\,,
\eqae 
where we have listed the types of SO(16) generators that can appear in the double-soft limit. 

\par
As a simple example, let us consider the case with $(m_1,m_2)=(8,0)$ in detail. In the double-soft limit, the amplitude takes the form, 
\beqa
\label{m80}
 \mathcal M_n^{(8_A, 0_B)}=\epsilon_{I_1\dots I_8}\sum_{a=3}^{n-1}\left(Z_a+ S_a\right)M_{n-2}+\mathcal O(\epsilon)\,,
\eeqa
where the operator $Z_a$ and $S_a$ are defined as
\beqa
\notag &&Z_a\equiv-\sum_{j=1,2}f_0(z_j)(\alpha_j\beta_j\eta_a^I-\alpha^2_j\eta_n^I)\frac{\partial}{\partial \eta_n^I}-4\sum_{j=1,2}f_0(z_j)\alpha_j \, , \cr
\notag && S_a\equiv\sum_{j=1,2}\left[\frac{f_0(z_j)}{\epsilon^2}+\frac{f_1(z_j)}{\epsilon}+f_2(z_j)
+\sum_{k=\lambda_a,\eta_n,\eta_a}f_0(z_j)O_{k}(z_j)\right]\,.
\eeqa
Here $I$ is the SU(8) index carried by the soft particle $\xi^{(8_A)}$\,. There are two ways to distribute $8$ $\eta_1$'s\,. One is to take 7 $\eta_1$'s from $\delta_\sigma$ and 1 $\eta_1$ from $\mathcal U$\,, which is of order $\epsilon$ and corresponds to the first term in $Z_a$\,. The other case is to extract $8$ $\eta_1$'s purely from $\delta_\sigma$. When one of the $8$ $\eta_1$'s in $\delta_\sigma$ is of order 
$\epsilon^2$ and the other $7$ $\eta_1$'s are $\epsilon^0$\,, we have a term of $\epsilon^0$ which results in the second term in $Z_a$\,. The other terms purely from $\delta_\sigma$ with different orders of $\epsilon$ give us $S_a M_{n-2}$.
The first term in $Z_a$ can be further simplified, because of momentum conservation and super-momentum conservation, 
\beqa
\notag 
-\sum_{j=1,2}f_0(z_j)(\alpha_j\beta_j\eta_q^I-\alpha^2_j\eta_n^I)\frac{\partial}{\partial \eta_n^I}&=&\sum_{j=1,2}\frac{\beta_j-\beta^*_j}{4\beta_j}\left(\frac{\partial}{\partial \eta_n^I}\eta_n^I\right) M_{n-2}+\mathcal O(\epsilon)\\
&=&-\frac{8 p_n\cdot p_2}{p_n\cdot (p_1+p_2)}M_{n-2}+\mathcal O(\epsilon)\,.
\eeqa
Similarly for the second term in $Z_a$, by momentum conservation,
\beqa
-4\sum_{a=3}^{n-1}\sum_{j=1,2}f_0(z_j)\alpha_j M_{n-2}=\frac{4 p_n\cdot p_2}{p_n\cdot (p_1+p_2)}M_{n-2}+\mathcal O(\epsilon)\,.
\eeqa
As a result, we have
\beqa
\mathcal M_n^{(8_A,0_B)}=\epsilon_{I_1\dots I_8}\left[-\frac{4 p_n\cdot p_2}{p_n\cdot (p_1+p_2)}+\sum_{a=3}^{n-1} S_a \right]M_{n-2}\,.
\eeqa
A similar calculation gives us the following result for $\mathcal M_n^{(0_B,8_A)}$\,, 
\beqa
\mathcal M_n^{(0_B,8_A)}=\epsilon_{I_1\dots I_8}\left[\sum_{a=3}^{n-1} S_a+\sum_{a=3}^{n-1}\frac{4 p_a\cdot p_2}{p_a\cdot (p_1+p_2)}-\sum_{a=3}^{n}\frac{p_a\cdot p_2}{p_a\cdot (p_1+p_2)}\eta_a^I\frac{\partial}{\partial\eta_a^I}\right]M_{n-2}\,.
\eeqa
By the antisymmetric extraction, the symmetric term $S_a M_{n-2}$ cancels out, and we are left with the U(1) generator of U(8),
\beqa
\mathcal M_n^{(8_A,0_B)}-\mathcal M_n^{(0_B,8_A)}=\epsilon_{I_1\dots I_8}\sum_{a=3}^{n}\frac{p_a\cdot p_2}{p_a\cdot (p_1+p_2)}\left(\eta_a^I\frac{\partial}{\partial\eta_a^I}-4\right)M_{n-2}\,.
\eeqa
Other R-symmetry operators with $m_A+m_B=8$ can be found by the same method. Here we simply list the results:   
\beqa
\mathcal M_n^{(6_A,2_B)}-\mathcal M_n^{(2_B,6_A)}&=&\sum_{a=3}^{n}\frac{p_a\cdot p_2}{p_a\cdot (p_1+p_2)}\times\\
\notag&&~~~\left[
4\epsilon_{I_1\dots I_6 I[J_1}\eta_a^I\frac{\partial}{\partial\eta_a^{J_2]}}
+\epsilon_{I_1\dots I_6J_1J_2}\left(\eta_a^I\frac{\partial}{\partial\eta_a^I}-2\right)\right]M_{n-2}\,,
\eeqa
\beqa
\mathcal M_n^{(4_A,4_B)}-\mathcal M_n^{(4_B,4_A)}&=&
\sum_{a=3}^{n}\frac{p_a\cdot p_2}{p_a\cdot (p_1+p_2)}\times\\
\notag&&~~~\left(8\epsilon_{I_1\dots I_4I[J_1J_2J_3}\eta_a^I\frac{\partial}{\partial\eta_a^{J_4]}}+\epsilon_{I_1\dots I_4J_1\dots J_4}\eta_a^I\frac{\partial}{\partial\eta_a^I}\right)M_{n-2}\,.
\eeqa
One can immediately see that after taking into account the explicit form of the SO(16) generators in eq.(\ref{Gen1}) and eq.(\ref{Gen2}), the above result can be rewritten as:
\beqa
\notag \mathcal M_n^{(8_A,0_B)}-\mathcal M_n^{(0_B,8_A)}&=&\displaystyle-\sum_{a=3}^{n}\frac{p_a\cdot(p_1- p_2)}{2p_a\cdot (p_1+p_2)}\left(R_a\right)M_{n-2}\\
\notag \mathcal M_n^{(6_A,2_B)}-\mathcal M_n^{(2_B,6_A)}&=&\displaystyle-\sum_{a=3}^{n}\frac{p_a\cdot(p_1- p_2)}{2p_a\cdot (p_1+p_2)}\times\\
\notag &&~~~~~\left[
4\epsilon_{I_1\dots I_6 I[J_1}(R_a)^I\,_{J_2]}+\frac{1}{2}\epsilon_{I_1\dots I_6J_1J_2}\left(R_a\right)\right]M_{n-2},\\
\mathcal M_n^{(4_A,4_B)}-\mathcal M_n^{(4_B,4_A)}&=&\displaystyle-
\sum_{a=3}^{n}\frac{p_a\cdot(p_1- p_2)}{2p_a\cdot (p_1+p_2)}\left[8\epsilon_{I_1\dots I_4I[J_1J_2J_3}(R_a)^I\,_{J_4]}\right]M_{n-2}\,.
\eeqa

Similarly we can consider the case with $m_A+m_B=6$ and $m_A+m_B=10$, the results are given by
\beqa
\notag 
\mathcal M_{n}^{(6_A,0_B)}-\mathcal M_{n}^{(0_B,6_A)}&=&\sum_{a=3}^{n}\frac{p_a\cdot(p_1- p_2)}{2p_a\cdot (p_1+p_2)}\left[\epsilon_{I_1\dots I_6 IJ}(R_a)^{IJ} \right]M_{n-2}\,,\\
\nonumber\mathcal M_{n}^{(4_A,2_B)}-\mathcal M_{n}^{(2_B,4_A)}
&=&-\sum_{a=3}^{n}\frac{p_a\cdot(p_1- p_2)}{2p_a\cdot (p_1+p_2)}\left[\epsilon_{I_1\dots I_4J_1J_2IJ}(R_a)^{IJ}\right]M_{n-2}\\
\notag \mathcal M_{n}^{(8_A,2_B)}-\mathcal M_{n}^{(2_B,8_A)}
&=&-\sum_{a=3}^{n}\frac{p_a\cdot(p_1- p_2)}{2p_a\cdot (p_1+p_2)}\left[-2\epsilon_{I_1\dots I_8}(R_a)_{IJ}\right]M_{n-2},\\ 
\mathcal M_{n}^{(6_A,4_B)}-\mathcal M_{n}^{(4_B,6_A)}&=&-\sum_{a=3}^{n}\frac{p_a\cdot(p_1- p_2)}{2p_a\cdot (p_1+p_2)}\left[6\epsilon_{I_1\dots I_6[J_1J_2}(R_a)_{J_3J_4]}\right]M_{n-2}\,.
\eeqa

Finally for the cases with $m_A+m_B>10$ or $m_A+m_B<6$, it turns out the amplitudes vanish in the limit. Thus, by studying the double-soft-scalar limits of the tree-level amplitudes, we have derived all non-trivial R-symmetry operators for three-dimensional $\mathcal N=16$ supergravity. Furthermore, we have reconstructed the algebra of $E_{8(8)}$ with part of the $SO(16)$ being non-linearly realised. In the next section, we will consider the fate of these double-soft limits at one-loop.

 \subsection{The double-soft limit: One Loop}
In previous section, we have derived the result of double-soft limits for tree-level amplitudes in $\mathcal N=16$ supergravity, in this section we will consider the double-soft limits at one-loop using generalized unitary cuts in three dimensions~\cite{unitary3D}. The integral representation of the three-dimensional theory can be deduced from its four-dimensional parent which is known to be expressible as linear combinations of scalar box integrals. The four-dimensional integral, upon dimensional reduction, can be viewed as the definition of the dimensionally regulated three-dimensional integral, which captures all subtleties related to $-2\epsilon$ dependent terms. In three-dimensions, a scalar box-integral can be written as a linear combination of scalar triangles up to $\mathcal{O}(\epsilon)$ terms, thus one-loop amplitudes for three-dimensional maximal supergravity do not contain rational terms, and can be expressed in terms of scalar triangle integrals only.

So as we discussed above that an $n$-point one-loop amplitude can be expressed in terms of scalar triangle integrals $I^{\rm tri}$ with certain coefficients, namely, 
\beq
M_n = \sum_i c_i \, I^{\rm tri}_i \, .
\eeq
The coefficients $c_i$ can be most easily determined by the triple-cut, and written as a product of three tree-level amplitudes with a Jacobian factor from the cuts, 
\beq
c_i = {1 \over \sqrt{K_A^2 K^2_B K^2_C} } \int  d^8 \eta_{\ell_1} d^8 \eta_{\ell_2} d^8 \eta_{\ell_3} 
M^{\rm tree}_1 M^{\rm tree}_2 M^{\rm tree}_3  \, ,
\eeq
where $\ell_i$'s are the cut propagators and $K_A, K_B, K_C$ are the external momenta of three corners in the cuts. First, it is straightforward to see that the coefficients $c_i$ vanish in the single-soft limit. In what follows, we will mostly study the double-soft limits on the coefficients. There are three different situations as shown below, \\
\begin{center}
$\begin{array}{ccc}
~~\includegraphics[scale=0.7]{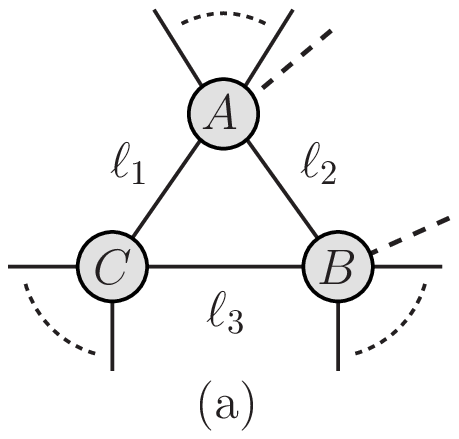}~~~~&~~~~\includegraphics[scale=0.7]{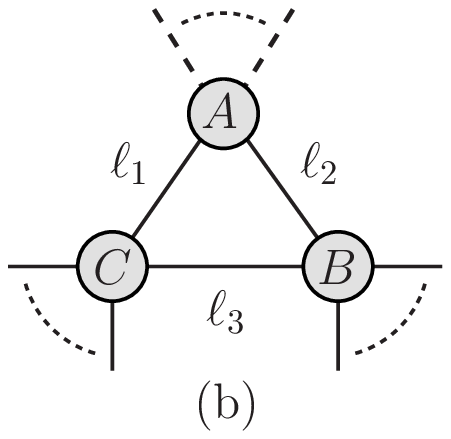}~~~~&~~~~\includegraphics[scale=0.7]{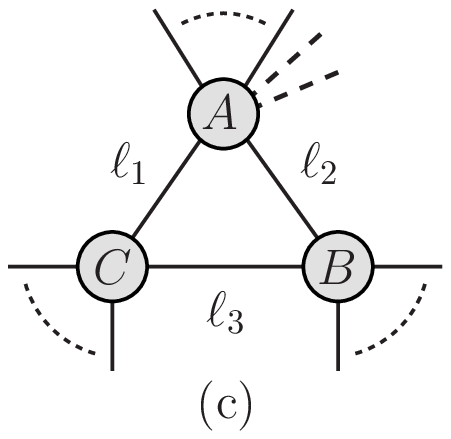}
\end{array}$\,,
\end{center}
where the dashed lines indicate the soft legs. It is easy to see that the first two kinds of diagrams, namely diagrams (a) and (b), vanish in the double-soft limits: the vanishing of diagram (a) is inherent from the result of the single-soft limit of tree-level amplitudes. One may worry about the degenerate cases where a soft leg is with only one hard leg at a corner, since the Jacobian factor $1/\sqrt{K_A^2 K^2_B K^2_C}$ is divergent as $1/\epsilon$ in this case, however the four-point amplitude goes as $\epsilon^2$ in the single-soft limit; as for diagram (b), we first note that $\ell_1, \ell_2$ are nothing but BCFW shifted soft legs, namely 
\beq
\lambda_{\ell_1} = c \lambda_p + s \lambda_q\, , \quad 
\lambda_{\ell_2} = c \lambda_q - s \lambda_p \, ,
\eeq
with $c^2+s^2=1$\,, and $p, q$ are the soft legs. Thus in the double-soft limits, we have not only $\lambda_p, \lambda_q \rightarrow 0$\,, but also $\lambda_{\ell_1}\,, \lambda_{\ell_2} \rightarrow 0$\,, namely all the external legs of the four-point amplitude become soft, and it is straightforward to check that the amplitude goes as $\epsilon^4$ in this case, while the cut Jacobian is divergent only as $1/\epsilon^2$.

Let us now consider the interesting case, namely diagram (c), where two soft legs (with some hard legs) are at the same corner of the triple cut. First we note that following diagrams lead to the same triangle integral in the soft limit,
\begin{center}
$\begin{array}{ccc}
~~\includegraphics[scale=0.7]{loopc}~~~~&~~~~\includegraphics[scale=0.7]{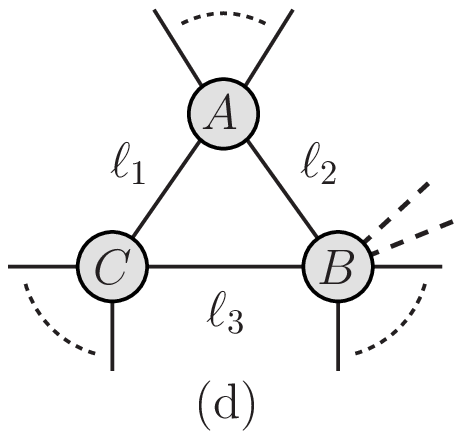}~~~~&~~~~\includegraphics[scale=0.7]{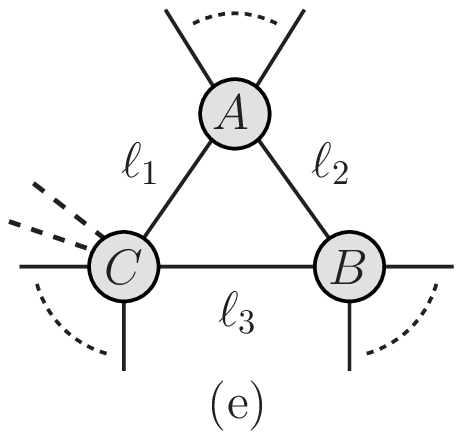}
\end{array}$\,.
\end{center}
So they should be combined together. Taking the diagram (c) as an example, the coefficient is given by the triple-cut,
\beq
C_{\rm (c)} =
{J_{\rm (c)} }\int d^8 \eta_{\ell_1} d^8 \eta_{\ell_2} d^8 \eta_{\ell_3}
M(A; p, q, \ell_1, -\ell_2) M(B; \ell_2, -\ell_3) M(C; \ell_3, -\ell_1)\,,
\eeq
where $A, B, C$ denote the hard legs at the three corners respectively, $p, q$ are the soft legs, and the Jacobian $J_{\rm (c)} = 1/\sqrt{(K_A+p+q)^2 K^2_B K^2_C}$\,. Using the result of the double-soft limit for the tree-level amplitudes, we have
\beq
C_{\rm (c)}  \rightarrow 
C^{(A)}_{\rm (c)} + C^{(\ell_1)}_{\rm (c)} + C^{(\ell_2)}_{\rm (c)} \, ,
\eeq
where each term is given by
\beqa
C^{(A)}_{\rm (c)} &=&{1 \over \sqrt{K^2_A K^2_B K^2_C} }  
\sum_{j \in A}  \mathcal{S}_j \left( \int  \prod^3_{i=1} d^8 \eta_{\ell_i}  M(A; \ell_1, -\ell_2) M(B; \ell_2, -\ell_3) M(C; \ell_3, -\ell_1)  \right)\,, \cr 
C^{(\ell_1)}_{\rm (c)} &=&{1 \over \sqrt{K^2_A K^2_B K^2_C} }  
\int  \prod^3_{i=1} d^8 \eta_{\ell_i}
\left(  \mathcal{S}_{\ell_1}   M(A; \ell_1, -\ell_2) \right) M(B; \ell_2, -\ell_3) M(C; \ell_3, -\ell_1)\,,  \cr
C^{(\ell_2)}_{\rm (c)} &=&{1 \over \sqrt{K^2_A K^2_B K^2_C} }  
\int  \prod^3_{i=1} d^8 \eta_{\ell_i}
\left(  \mathcal{S}_{\ell_2}   M(A; \ell_1, -\ell_2) \right) M(B; \ell_2, -\ell_3) M(C; \ell_3, -\ell_1)  \, .
\eeqa
where $\mathcal{S}_i$ are the double-soft factors, which may be proportional to one of the generators we defined in previous section, 
\beqa 
 R_i^{IJ} \, , \quad  (R_i)_{IJ}  \, , 
\quad (R_i)^I_{\;\,J} \, , \quad R_i  \, .
\eeqa
Similarly one can obtain the results for diagram (d) and (e), summing over all three diagrams, from the summation over the external legs, we find
\beqa
&&C^{(A)}_{\rm (c)} + C^{(B)}_{\rm (d)} + C^{(C)}_{\rm (e)}
=
{1 \over \sqrt{K^2_A K^2_B K^2_C} }  \cr
&&~~~~\times 
\sum_{j \in \{A,B,C \} }  \mathcal{S}_j \left( \int  \prod^3_{i=1} d^8 \eta_{\ell_i}  M(A; \ell_1, -\ell_2) M(B; \ell_2, -\ell_3) M(C; \ell_3, -\ell_1)  \right)\,,
\eeqa
which is precisely the result of the double-soft limits at one-loop. So we need to prove the contributions from internal lines vanish. That is indeed true, they all cancel in pairs, as we will prove 
\beq
C^{(\ell_1)}_{\rm (c)} + C^{(\ell_1)}_{\rm (e)} =0\, , \quad
C^{(\ell_2)}_{\rm (c)} + C^{(\ell_2)}_{\rm (d)} =0\, , \quad
C^{(\ell_3)}_{\rm (d)} + C^{(\ell_3)}_{\rm (e)} =0 \, .
\eeq
Let us take $C^{(\ell_1)}_{\rm (c)} + C^{(\ell_1)}_{\rm (e)}$ as an example. First for the case when the soft factor is proportional to $ R_{\ell_1}^{IJ} = \eta_{\ell_1}^I \eta_{\ell_1}^J$, we have
\beqa
C^{(\ell_1)}_{\rm (c)} + C^{(\ell_1)}_{\rm (e)} &\sim &
\Big[ \eta^I_{\ell_1} \eta^J_{\ell_1}   M(A; \ell_1, -\ell_2) M(B; \ell_2, -\ell_3) M(C; \ell_3, -\ell_1)
\cr
&&+ \eta^I_{-\ell_1} \eta^J_{-\ell_1} M(A; \ell_1, -\ell_2)  M(B; \ell_2, -\ell_3)    M(C; \ell_3, -\ell_1) \Big] \, .
\eeqa
Now, using the fact that $\eta^P_{-\ell_1} = i \eta^P_{\ell_1}$, we find the above two terms cancel out precisely. The same argument applies to $(R_{\ell_1})_{IJ}$. Let us now consider $(R_{\ell_1})^I~_J$, and we have, 
\beqa
C^{(\ell_1)}_{\rm (c)} + C^{(\ell_1)}_{\rm (e)}
&\sim &
\int  \prod^3_{i=1} d^8 \eta_{\ell_i}
\Big[ \eta^I_{\ell_1} \left(  (R_{\ell_1})^I~_J   M(A; \ell_1, -\ell_2) \right) M(B; \ell_2, -\ell_3) M(C; \ell_3, -\ell_1)
\cr
&&+  M(A; \ell_1, -\ell_2)  M(B; \ell_2, -\ell_3)  \left(  (R_{\ell_1})^I~_J  M(C; \ell_3, -\ell_1) \right)  \Big]
\cr
&=&
\int  \prod^3_{i=1} d^8 \eta_{\ell_i}
 (R_{\ell_1})^I~_J  \Big[   M(A; \ell_1, -\ell_2)  M(B; \ell_2, -\ell_3) M(C; \ell_3, -\ell_1) \Big] \, .
\eeqa
When $I \neq J$, $(R_{\ell_1})^I~_J = \eta^I_{\ell_1} \partial_{\eta^J_{\ell_1}}$, and the above result is a total derivative, which then vanishes trivially under the fermionic integration. Whereas when $I = J$, the vanishing of the above result can be seen by using the identity $\int d\eta\, \eta\frac{\partial}{\partial \eta} \,\ast=\int d\eta \,\ast$. Finally, we consider when the soft factor is proportional to the U$(1)$ generator $R_{\ell_1} $, we have\footnote{This type of argument was first realized in~\cite{BTWsoft} for $\mathcal{N}=8$ supergravity in four dimensions.}
\beqa
C^{(\ell_1)}_{\rm (c)} + C^{(\ell_1)}_{\rm (e)}
&\sim& 
\int  \prod^3_{i=1} d^8 \eta_{\ell_i}
\left( \sum^8_{K=1} \eta^K_{\ell_1} \partial_{\eta^K_{\ell_1}} -8 \right)  \nonumber
\\
&&~~~~~~~~~~~~~~\times\Big[   M(A; \ell_1, -\ell_2)  M(B; \ell_2, -\ell_3) M(C; \ell_3, -\ell_1) \Big] \, .
\eeqa
Due to the same identity, $\int d\eta\, \eta\frac{\partial}{\partial \eta} \,\ast=\int d\eta \,\ast$, we find that the above sum vanishes. Thus we have proved all possible soft factors in $\mathcal{N}=16$ supergravity do not receive any one-loop correction. 

It is easy to see that the above discussion, namely the cancellation between internal cut propagators, is actually valid for any multiple cuts, if the lower-point (loop) amplitudes entering in the cuts satisfy the double-soft theorems. This observation as well as the explicit calculation for one-loop amplitudes we have done strongly suggest that any higher-loop amplitudes should behave in the same way as the tree-level amplitudes in the double-soft-scalar limits, namely duality symmetries do not receive any loop corrections. 
\section{Implications for counter terms in $\mathcal{N}=16$ SUGRA} \label{section:counterterm}
In the previous sections, we have demonstrated both at tree- and one-loop level, in the single-soft-scalar limit the $n$-point scattering amplitude vanishes while the double-soft-scalar limit is given by an SO(16) rotation on the $M_{n-2}$. The same constraint applies to matrix-elements generated by possible counter terms, and thus provides an on-shell check on whether or not potential counter terms respect the duality symmetry of the theory. This line of approach has been extensively pursued for the four-dimensional $\mathcal{N}=8$ theory~\cite{LanceR4, ElvangSUSY, ElvangR4, ElvangFull}. Given that all degrees of freedom in three-dimensional supergravity are subject to duality constraints, we expect that the constraint on counter terms are much more stringent compared to its four-dimensional counter part. In particular, in three dimensions, all bosonic soft-limits must vanish. Furthermore, using supersymmetric Ward identities~\cite{SUSYWard}, one can deduce that all single-soft-fermion limits should vanish as well, since:
\eq
\langle [Q,\phi_1\cdots]\rangle=0= \langle q1\rangle\langle \psi_1\cdots\rangle+\langle \phi_1[Q,\cdots]\rangle\,,
\eqe 
where $|q\rangle$ is an auxiliary spinor and $\cdots$ represent a collection of fields. The single-soft-scalar limit vanishes as $\mathcal{O}(\epsilon^2)$, thus in order for the RHS to vanish, at the very least $\langle \psi_1\cdots\rangle$ has to be of $\mathcal{O}(\epsilon)$ in the single-soft limit. Thus remarkably, supersymmetry combined with the duality symmetry implies that all single-soft limits must vanish for three-dimensional supergravity!  
 
At four points, $\mathcal{N}=16$ supersymmetry requires the matrix element of any four-point operator to be of the form 
\eq
\mathcal{L}_4=\delta^{16}(Q)f(s,t,u)\,,
\eqe 
where the function $f(s,t,u)$ is a polynomial and symmetric in $s,t,u$. This matrix-element is invariant under the full SO(16) R-symmetry. The fact that it is invariant under $\mathcal{R}^{ I}\,_{ J}$\footnote{Here, we use $\mathcal{R}$ to denote all R-symmetry generators to distinguish them from the field strength $R$.} is trivial, where as for $\mathcal{R}^{ IJ}$ let us choose $I=7,J=8$ in $\mathcal{R}^{IJ}\delta^{16}(Q)=\sum_{i=1}^4\eta^{ I}_i\eta_i^{ J}\delta^{16}(Q)$\,, and consider the component $\eta^7_1\eta^7_2\eta^7_3\eta^8_2\eta^8_3\eta^8_4$\,. It is given by:
\eqa
\int d\eta^7_1d\eta^7_2d\eta^7_3d\eta^8_2d\eta^8_3d\eta^8_4\;\sum_{i=1}^4\eta^{\rm I}_i\eta_i^{\rm J}\delta^{16}(Q)\sim (\langle 12\rangle\langle23\rangle+\langle 14\rangle\langle43\rangle)=0\,.
\eqae
Similar analysis applies to $\mathcal{R}_{IJ}$\,. Thus at four-point we can have:
\eq\label{4ptResult}
\delta^{16}(Q)\,,\;\;\delta^{16}(Q)(s^2+t^2+u^2)\,, \;\;\delta^{16}(Q)(s^3+t^3+u^3)\,,\; \;\delta^{16}(Q)(s^2+t^2+u^2)^2\,.
\eqe
These counter term elements can be viewed as the descendant of four-dimensional elements generated from $R^4$, $D^4R^4$, $D^6R^4$ and $D^8R^4$ respectively. In three dimensions, an operator $\mathcal{O}$ of mass-dimension $m$ corresponds to $m=3+(L-1)$ loops, and hence the above would correspond to possible counter terms for $6-, 10-, 12$- and $14$-loop divergence respectively. Note that leading ultraviolet divergences\footnote{By leading we are referring to the first place where ultra-violet divergence is present} is automatically ruled out for odd-loops.

\subsection{Matrix elements from dimensional reduction}
To test the validity of these operators, we need to see if the corresponding six-point matrix elements satisfy the required soft-behavior. Given that the four-point matrix elements discussed in eq.(\ref{4ptResult}) are direct descendants of their four-dimensional counter parts, we will test whether their six-point descendants satisfy the duality symmetry in three dimensions. In other words, \textit{we will consider whether or not direct dimensional reduction of $R^4$, $D^4R^4$, $D^6R^4$ and $D^8R^4$ to three-dimensions yield valid counter terms.} Here, by dimensional reduction, we mean to substitute three-dimensional kinematics to the four-dimensional amplitudes. Recall that the SO(16) scalars are organized as a representation of SU(8):
$$
\begin{tabular}{c|c|c|c|c}
\hline
  1 &  $\eta^2$ & $\eta^4$ & $\eta^6$ & $\eta^8$\\ \hline
    $\xi$ & $\xi_{ IJ}$ & $\xi_{ IJKL}$& $\overline{\xi}^{IJ}$& $\overline{\xi}$\\
\hline
 1 & 28 & 70 & 28 & 1\\
 \hline
\end{tabular}\,\;.
$$
It is then straightforward to identify the states between three and the four-dimensional counter part. In particular, using the notation in ~\cite{ElvangSUSY} we can identify
\eq
h^+ \leftrightarrow \xi,\quad v^{+ IJ} \leftrightarrow \xi_{ IJ},\quad \varphi^{ IJKL} \leftrightarrow \xi_{ IJKL},\quad v^{\rm -}_{ IJ} \leftrightarrow \overline{\xi}^{IJ},\quad h^- \leftrightarrow \bar{\xi}\,.
\eqe
In~\cite{ElvangR4, ElvangFull} it was shown that the $R^4$, $D^4R^4$, and $D^6R^4$ operators have non-vanishing single-soft limit. In the limit the matrix element becomes proportional to a local quantity that does not vanish in three-dimensional kinematics.\footnote{More precisely, for the amplitude $M_6(- - + + \varphi \bar{\varphi})$ of $R^4$ $D^4R^4$, $D^6R^4$ behaves in the soft limit as $\langle 12\rangle^4[34]^4$, $\langle 12\rangle^4[34]^4(\sum_{i<j\leq4}s^2_{ij})$, and $\langle 12\rangle^4[34]^4(\sum_{i<j\leq4}s^3_{ij})$ respectively. } Thus such matrix elements also yield incorrect single-soft limit in three-dimensions, and one can rule them out as possible counter terms.

As a consistency, we would like to show that the finite single-soft-limit for the matrix elements of $R^4, D^4R^4, D^6R^4$ are in fact universal: i.e. the same is true for the three-dimensional reduction of soft gravitons and vectors. This is non-trivial since in four dimensions the single-soft limit of a graviton is divergent: 
\eq\label{SoftExpand}
\mathcal{M}_{n+1}(1,\cdots,n,s)=\frac{1}{\epsilon^2}\mathcal{S}_G^{(0)}\mathcal{M}_{n}(1,\cdots,n)+\mathcal{S}_G^{(1)}\mathcal{M}_{n}(1,\cdots,n)+\cdots\,.
\eqe
Since it was shown in~\cite{HHMW} that the above graviton soft-theorem is universal for any effective theory for gravity, it is applicable to our counter term matrix element. Taking $h^-$ to be soft, $\mathcal{S}_G^{(0)}$ is given by:
\eq
\mathcal{S}_G^{(0)}=\sum_{a=1}^{n}\left( \frac{[ \mu a ]}{[ \mu s ]}\langle sa \rangle\right)^2\frac{1}{[ as ] \langle sa \rangle}\quad \xrightarrow[~\mbox{ kinematics}~]{\mbox{ 3D}}\quad \sum_{a=1}^{n} \frac{\langle \mu a \rangle^2}{\langle \mu s\rangle^2}=-1\,.
\eqe
Since $\mathcal{S}_G^{(0)}$ is a constant in three-dimensional kinematics, its contribution is simply $-M_{\rm odd}$. As the leading contribution is finite, it is at the same order as with $\mathcal{S}_G^{(1)}$, and must be combined. Next, let's consider the single-soft limit of the vectors. Using the supersymmetric soft theorems, one can show that 
\eq
\mathcal{M}_{n+1}(1,\cdots,n,s)|_{v_{IJ}}=\mathcal{S}^{(0)}_{v_{IJ}}\mathcal{M}_{n}\,,
\eqe
where $\mathcal{S}^{(0)}_{v^{IJ}}$ is the soft operator due to a soft vector $v^{IJ}$ and is given by:
\beqa
\nonumber && \mathcal{S}^{(0)}_{v_{IJ}}=
\sum^n_{a=1} \langle s a\rangle[is]\frac{[ na ]^2}{[ ns ]^2 [ sa ]^2} 
\left( { [ns] \over [an]} \eta^I_i  + { [s a] \over [an]} \eta^I_n \right)
\left( { [ns] \over [an]} \eta^J_i  + { [s a] \over [an]} \eta^J_n \right) \\
&&
~~~~~~~~~~~~~~~~~~~\xrightarrow[~\mbox{ kinematics}~]{\mbox{ 3D}}
\quad 
\sum^n_{a=1} 
\left(  \eta^I_a  + {\langle s a \rangle \over \langle ns \rangle } \eta^I_n \right)
\left(  \eta^J_a  + { \langle s a \rangle \over \langle ns \rangle } \eta^J_n \right)\,.
\eeqa
Again $\mathcal{S}^{(0)}_{v^{IJ}}$ is finite. Thus just as with the soft-graviton case, the vector-soft-limit in three-dimensional kinematics is proportional to $\mathcal{M}_{\rm odd}$\,. 

To obtain the five-point matrix elements of $R^4$, $D^4R^4$ and $D^6R^4$, one can consider the $\alpha'$ expansion of the string theory five-point tree-level amplitude. This is because string theory can be viewed as an effective field theory with higher dimensional operators whose cut off scale is $\alpha'$. In particular, the tree-level effective action for closed strings are given by $S_{eff}$:
\eq
S_{eff}=S_{SG}-2\alpha'^{3}\zeta(3)e^{-6\phi}R^4-\alpha'^{5}\zeta(5)e^{-10\phi}D^4R^4+\cdots\,,
\eqe
where $S_{SG}$ is the supergravity action and $\phi$ is the dilaton. As one can see, different order of $\alpha'$ in the expansion of the closed string amplitude can be matched with the matrix elements generated by different higher dimensional operators in $S_{eff}$.

There is no need to worry about possible contamination by other candidate operators at five-point, such as $D^2R^5$ and $D^4R^5$, since they are already ruled out by linearly realized four-dimensional supersymmetry~\cite{ElvangSUSY}. Explicit evaluation of the five-point closed string amplitude reveals that it is indeed non-zero at $\alpha'^4$ and $\alpha'^6$, while $\mathcal{S}_G^{(0)}$ acting on it is also non-vanishing after dimension reduction, and the two do not cancel. This confirms that gravitons, vectors, and scalars single soft-limits are finite and non-vanishing. In contrast, all single-soft limits vanish since  $\mathcal{M}_{\rm odd}=0$.

The fact that all bosonic soft limits vanish in three dimensions is quite remarkable. This is to be compared with Yang-Mills, where one has:
\eq
\mathcal{A}_{n+1}(1,\cdots,n,s)=\mathcal{S}^{(0)} \mathcal{A}_{n}(1,\cdots,n)+\mathcal{S}^1\mathcal{A}_{n}(1,\cdots,n)+\cdots\,,
\eqe
with $\mathcal{S}^{(0)}$ given as:
\eq
\mathcal{S}^{(0)} =\frac{\langle 1\,s\rangle}{\langle 1\,n\rangle \langle n\,s\rangle}\,.
\eqe
This does not change in three-dimensional kinematics, and thus Yang-Mills is still soft-divergent in $D=3$.

Finally, we consider the $D^8R^4$ operator, which has a vanishing single-soft limit in four dimensions. The enlarged duality group now requires that the soft-graviton limit also has to vanish in three-dimensional kinematics. As discussed above it is proportional to the five-point amplitude. To extract the five-point matrix element of $D^8R^4$ operator, we have evaluated the $\alpha'^7$ term in the five-point amplitude, and the application of $\mathcal{S}_G^{(0)}$ on it. The two are non-zero and they do not cancel. Thus the six-point amplitude of $D^8R^4$ is non-vanishing in the single soft limit, ruling it out as a potential counter term. Again, our five-point analysis is not polluted by  $D^6R^5$ since it is ruled out in four dimensions. In conclusion, the dimensional reduction of the four-dimensional $D^8R^4$ operator violates the single-soft limit in three dimensions.
\subsection{SO(16) and SU(8)}
The four-dimensional $D^{2k}R^4$ matrix elements discussed above were extracted from the $\alpha'$ expansion of string amplitudes in~\cite{ElvangR4, ElvangFull}. These operators appear in the string theory effective action dressed with factors of dilatons, and thus special care is required to insure the projection of these unwanted states. In practice, this is done by an averaging over all inequivalent embedding of the SU(4)$\times$SU(4) symmetry of string theory into the SU(8) R-symmetry of $\mathcal{N}=8$ supergravity.

In our case, the direct dimensional reduction of four-dimensional matrix elements are only guaranteed to have SU(8) invariance, and thus one might wonder if a similar SO(16) ``averaging" is needed. While a priori there is no need to do so, as there are no extra contributions in four dimensions for which we might want to project out, it is useful to consider this as a trick to obtain manifest SO(16) invariant matrix-elements that potentially can have vanishing single-soft limits. Here we will demonstrate that for $R^4$ and $D^4R^4$, this is not possible. We first introduce the correct SO(16) averaging procedure.

The scalars of the theory carry chiral-spinor indices of SO(16), they carry indices $A=1,\cdots,64$ which are complex spinor indices. For SO(16) invariants they are paired with their complex conjugate. In the SU(8) representation, these 64 pairs of complex scalars are cast into $1\oplus 28\oplus 35$ pairs, $(\Phi, \bar{\Phi})$, $(\Phi^{IJ},\bar{\Phi}_{IJ})$ $(\Phi^{IJKL},\bar{\Phi}_{IJKL})$. For an SO(16) invariant counter term amplitude, one can consider the following linear combination:
\eq \label{average}
\frac{1}{64}M_6(\Phi, \bar{\Phi},\cdots)+\frac{28}{64}M_6(\Phi^{IJ},\bar{\Phi}_{IJ},\cdots)+\frac{35}{64}M_6(\Phi^{IJKL},\bar{\Phi}_{IJKL},\cdots)\,.
\eqe
In other words, the non-vanishing single-soft limit of the scalar amplitudes in four dimensions can potentially be combined with the single-soft limit of the gravitons and the vectors to obtain vanishing result. 

To be precise, let us consider the six-point amplitudes with specific helicity configurations, for instance, 
\eq
\frac{1}{64}M_6(\Phi, \bar{\Phi},+, +, -, -)+\frac{28}{64}M_6(\Phi^{IJ},\bar{\Phi}_{IJ},+, +, -, -)+\frac{35}{64}M_6(\Phi^{IJKL},\bar{\Phi}_{IJKL},+, +, -, -)\,.
\eqe
Firstly, it is known from the result of~\cite{ElvangR4, ElvangFull}, after reducing to three dimensions, we have
\eqa
&&\hspace{-1cm} M_6(\Phi^{IJKL},\bar{\Phi}_{IJKL},+, +, -, -) \quad 
\xrightarrow[~~]{~p_1\rightarrow 0~}
\cr
&&~~~~~~~~~~~ -2 \zeta_3 \alpha'^3 \langle 34\rangle^4 \langle 56\rangle^4 \left( 1 + a_1 \alpha'^5 \!\!\! \sum_{3\leq i<j\leq 6}s^2_{ij} + a_2 \alpha'^6 \!\!\! \sum_{3\leq i<j\leq 6 }s^3_{ij} + \ldots \right) \, ,
\eqae
where the coefficients $a_i$ are numbers with certain transcendentalities, whose precise results are not important for our discussion here. 
Whereas the single-soft limit of $M_6(\Phi, \bar{\Phi},+, +, -, -)$ and  $M_6(\Phi^{IJ},\bar{\Phi}_{IJ},+, +, -, -)$ can be determined by the soft graviton and graviphoton in four dimensions as we discussed previously. The results are simply given by certain soft factors acting the five-point amplitude. We have checked numerically that the combination of these two contributions have different kinematics dependence comparing to those from the soft limit of $M_6(\Phi^{IJKL},\bar{\Phi}_{IJKL},+, +, -, -)$ at each order. So we find the SO(16) averaging, eq.(\ref{average}), cannot save the non-vanishing results of single-soft limit for $R^4$ and $D^4R^4$. For $D^6R^4$, we need to carefully perform the averaging on the string theory element, as well as project out the SO(16) ``averaged" contribution of $R^4\frac{1}{\Box}R^4$. We leave this for future work.

\section{Conclusions} \label{section:conclusion}

In this paper, we investigate the action of the duality symmetries on the scattering amplitudes for a wide class of supergravity theories, including $\mathcal{N}=4,5,6$ supergravity in four dimensions at tree level, as well as $\mathcal{N}=16$ supergravity in three dimensions both at tree- and one-loop level. Not surprisingly, $\mathcal{N}=4,5,6$ supergravity can be studied from the better-understood maximal supergravity theory via susy reduction. However, the isotropy group of non-maximal supergravity theories contain a U$(1)$ factor, unlike their parent $\mathcal{N}=8$ theory. To expose this factor in the double-soft-scalar limit, the two scalars have to form a singlet, and the amplitude is divergent due to soft graviton exchanges. To extract the relevant information for this subtle U$(1)$ factor, we introduce the anti-symmetrised extraction procedure, which removes the singular part in the amplitude. This leaves behind a finite soft-factor that is proportional to the U$(1)$ generator of the isotropy group acting on a lower-point amplitude. Using this prescription we have successfully extracted the duality algebra for $4\leq\mathcal{N}<8$ supergravity theories. 

We then study the maximal supersymmetric gravity theory in three dimensions, namely $\mathcal{N}=16$ supergravity. As the on-shell variables only form a linearly representation of U(8)$\subset$ SO(16), the isotropy group contains pieces that are non-linearly realised along with a U$(1)$ factor. With the help of BCFW recursion relations in 3D as well as the anti-symmetrised extraction, for tree-level amplitudes we derive the results of double-soft limits for all scalar species. We then extend this result to one-loop amplitudes in the theory. We apply the generalized unitarity in 3D to express the one-loop amplitudes in terms of triangle integrals, whose coefficients are determined by the triple-cut. Since tree-level amplitudes enter in the cuts, the tree-level soft limits can be utilized directly, and we proved that all one-loop amplitudes satisfy the same soft theorems as the tree-level amplitudes. Unlike the case of $\mathcal{N}=8$ supergravity in 4D~\cite{BTWsoft}, where one needs to worry about discontinuities of box integrals in 4D, here such complication does not arise. That is because that only amplitudes with even number of external legs are non-vanishing, and thus a massive corner of a triangle integral can never suddenly become massless because of the soft limits. 

It would certainly be of great interest to prove (or disprove) that the soft theorems in $\mathcal N=16$ supergravity are tree-level exact, namely that they will not receive any loop corrections. First of all, one should expect that amplitudes at any loop order vanish in the single-soft limit, since the worst that loop-corrections can do is to generate some logarithmic singularities, which would not change the vanishing of the amplitudes, recall that the tree-level amplitudes vanish as $\epsilon^2$. As for the double-soft limits, the proof we used for one-loop amplitudes is actually quite general, and should be applicable to a general class of multiple-particle cuts relevant for higher-loop amplitudes. It thus already provides evidence that higher-loop amplitudes satisfy the same soft theorems as tree-level amplitudes. We leave the detailed investigation on higher-loop amplitudes for the future work. 

We also point out the curious vanishing single-soft fermion limit of $\mathcal{N}=16$ supergravity, which is implied by supersymmetric Ward identities. It will be interesting to see whether or not such soft behavior implies some kind of breaking of hidden fermionic symmetries. A preliminary step towards this direction would be the study of double-soft limits\cite{US}. Note that theories for which BCFW is applicable, such as ABJM and $\mathcal{N}=16$ supergravity, it is straightforward to deduce the single-soft limit. At four-points, as one takes $\lambda_1\rightarrow \epsilon \lambda_1$, momentum conservation ensures that all spinor inner brackets are of order $\epsilon$. Thus a four-point amplitude with mass dimension $M$ will behave in the single soft-limit as $\epsilon^{2M}$. For  ABJM and $\mathcal{N}=16$ supergravity, $M=1$ and $M=2$ respectively. 

To use the recursion, one has to consider the single-soft behavior of the function $H(z_1,z_2)$ in eq.(\ref{HDef}), where $(z_1,z_2)$ are solutions for a  four-point BCFW channel. It is straightforward to see that the function has an $1/\epsilon$ singularity. Through the recursion one sees would conclude that the single soft-limit of ABJM and $\mathcal{N}=16$ supergravity behaves as $\epsilon^0$ and $\epsilon^1$ respectively. However, it is easy to see that this is not correct. Changing $\epsilon\rightarrow -\epsilon$, is equivalent to performing a $Z_2$ flip on the spinor of the soft leg. Thus if the soft leg is a scalar (fermion), then the amplitude must be an even (odd) function of $\epsilon$. Thus due to the little group constraint, the leading soft-behavior might cancel exposing even more milder soft-limits. We find: 
\eqa
{\rm (ABJM)}:&& A(\phi_1\cdots)|_{\lambda_1\rightarrow \epsilon\lambda_1}\sim \epsilon^0\,, \quad A(\psi_1\cdots)|_{\lambda_1\rightarrow \epsilon\lambda_1}\sim \epsilon^1\,,\\
\mathcal{N}=16\;{\rm (SUGRA)}:&& M(\phi_1\cdots)|_{\lambda_1\rightarrow \epsilon\lambda_1}\sim \epsilon^2\,, \quad M(\chi_1\cdots)|_{\lambda_1\rightarrow \epsilon\lambda_1}\sim \epsilon^1\,.
\eqae
Thus we recover the result that the single-soft-scalar limit of supergravity amplitudes vanishes as $\epsilon^2$, while fermions vanish as $\epsilon$. Quite interestingly single fermions also vanish in ABJM theory.

Finally, along the line of the works in~\cite{ElvangFull, LanceR4, ElvangSUSY}, duality symmetry in supergravity plays important roles to constrain the potential UV counter terms. We apply this idea to study briefly the counter terms for $\mathcal N=16$ supergravity in 3D. The fact that the larger $E_{8(8)}$ imposes stronger constraint is reflected in the violation of the single-soft scalar limit for the dimensional reduction of matrix elements for $R^4$, $D^4R^4$, $D^6R^4$ and $D^8R^4$. Thus under the assumption that three-dimensional counter terms can be obtained by the direct reduction of four-dimensional ones, UV divergences are ruled out at $14$-loops for maximal supergravity in three-dimensions. Since there can not be any leading UV divergences at odd-loops, this implies the next possible divergence is at $16$-loops. We also consider possible SO(16) projections, and show that it cannot resolve the non-vanishing single-soft-limit for $R^4$ and $D^4R^4$ matrix elements. We leave the analysis for $D^6R^4$ and $D^8R^4$ for future investigations.

\section{Acknowledgements}
We would like to thank Henriette Elvang and Stephan Stieberger for their generosity in sharing their numerical codes that was the basis for their analysis in ~\cite{ElvangFull}. We also would like to thank Massimo Bianchi, Paul Howe for useful comments, C.W. would also like to thank Andreas Brandhuber and Gabriele Travaglini for the collaboration at the early stage of this work.

\end{document}